\documentclass{article}
\usepackage[hyphens]{url}
\usepackage[]{hyperref}

\usepackage{amsmath,amssymb,amsfonts}
\usepackage{algorithm}
\usepackage{algorithmic}
\usepackage{graphicx}
\usepackage{textcomp}
\usepackage{xcolor}
\usepackage{xspace}
\usepackage{mdframed}
\usepackage{tabularx}
\usepackage{tikz}
\usepackage{enumitem}
\usepackage{array}
\usepackage{color}
\usepackage{soul}
\usepackage{subcaption}
\usepackage{makecell}
\usepackage{multirow}
\usepackage{booktabs}
\usepackage{tablefootnote}
\usepackage{float}
\usepackage{colortbl}
\usepackage{microtype}
\usepackage{enumitem}

\usepackage{hyperref}

\definecolor{green0}{RGB}{243,250,248}
\definecolor{green1}{RGB}{233,244,235}
\definecolor{green2}{RGB}{223,238,222}
\definecolor{green3}{RGB}{213,232,209}
\definecolor{green4}{RGB}{203,226,196}
\definecolor{green5}{RGB}{193,220,183}
\definecolor{green6}{RGB}{183,214,170}
\definecolor{green7}{RGB}{173,208,157}
\definecolor{green8}{RGB}{163,202,144}
\definecolor{green9}{RGB}{153,196,131}

\newcommand{\ignore}[1]{}
\newcommand{\ie}{\textit{i.e.},\xspace}
\newcommand{\eg}{\textit{e.g.},\xspace}

\newcommand{\pgheading}[1]{\indent \textbf{#1.}}

\newcommand{\sys}{\textsc{TeleRAG}\xspace}

\newcommand{\lookah}{lookahead retrieval\xspace}
\newcommand{\nprobe}{\texttt{nprobe}\xspace}
\newcommand{\desktop}{\texttt{Desktop}\xspace}
\newcommand{\serverone}{\texttt{Server1}\xspace}
\newcommand{\servertwo}{\texttt{Server2}\xspace}
\newcommand{\llama}{Llama-3-8B\xspace}
\newcommand{\llamasmall}{Llama-3.2-3B\xspace}

\newcommand{\mistral}{Mistral-Small-22B\xspace}
\newcommand{\qin}{$q_\text{in}$\xspace}
\newcommand{\qout}{$q_\text{out}$\xspace}
\newcommand{\Cin}{$C_\text{in}$\xspace}
\newcommand{\Cout}{$C_\text{out}$\xspace}
\newcommand{\Coverlap}{$C_\text{overlap}$\xspace}
\newcommand{\Cmiss}{$C_\text{miss}$\xspace}
\newcommand{\npipeline}{six\xspace}

\newcommand{\gpusmall}{RTX4090\xspace}
\newcommand{\gpularge}{H100\xspace}
\newcommand{\gpuhuge}{H200\xspace}

\definecolor{shepherdpurple}{RGB}{102,0,153}
\newcommand{\cameraready}[1]{#1} 
\newcommand{\camerareadytwo}[1]{#1} 

\newcommand{\displayappendix}[2]{Appendix~{#2}}
\newcommand{\displayappendixref}[2]{{#2}}

\gdef\isaccepted{1}
\usepackage[accepted]{mlsys2026}


\mlsystitlerunning{\sys{}: Efficient Retrieval-Augmented Generation Inference with Lookahead Retrieval}

\AddToShipoutPictureBG*{
  \AtPageUpperLeft{
    \hspace*{\paperwidth}
    \raisebox{-68pt}{
      \llap{
        \href{https://www.acm.org/publications/policies/artifact-review-and-badging-current}{
          \includegraphics[height=65pt]{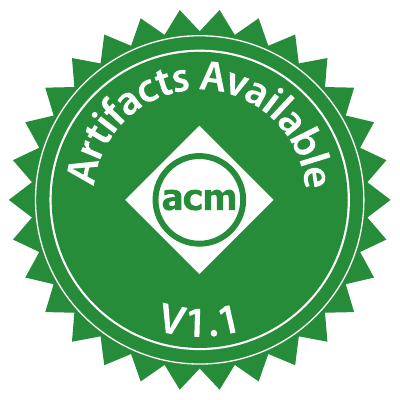}}
        \hspace{1pt}
        \href{https://www.acm.org/publications/policies/artifact-review-and-badging-current}{
          \includegraphics[height=65pt]{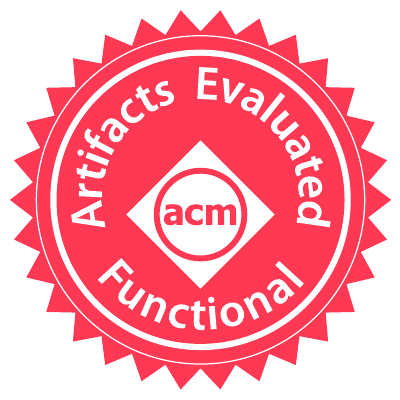}}
        \hspace{1pt}
        \href{https://www.acm.org/publications/policies/artifact-review-and-badging-current}{
          \includegraphics[height=65pt]{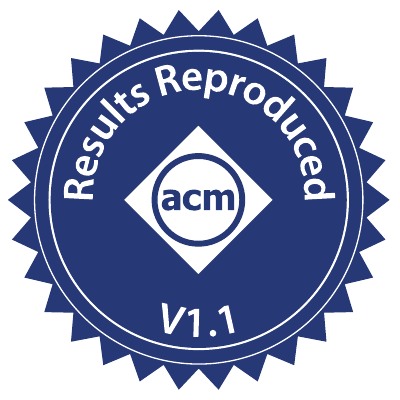}}
        \hspace{80pt}
      }
    }
  }
}

\begin{document}

\twocolumn[
\mlsystitle{\sys{}: Efficient Retrieval-Augmented Generation \\ 
Inference with Lookahead Retrieval}



\mlsyssetsymbol{equal}{*}
\mlsyssetsymbol{intern}{\textdagger}

\begin{mlsysauthorlist}
\mlsysauthor{Chien-Yu Lin}{equal,uw}
\mlsysauthor{Keisuke Kamahori}{equal,uw}
\mlsysauthor{Yiyu Liu}{equal,intern,harvard}
\mlsysauthor{Xiaoxiang Shi}{intern,sjtu}
\mlsysauthor{Madhav Kashyap}{uw}
\mlsysauthor{Yile Gu}{uw}
\mlsysauthor{Rulin Shao}{uw}
\mlsysauthor{Zihao Ye}{uw}
\mlsysauthor{Kan Zhu}{uw}
\mlsysauthor{Rohan Kadekodi}{uw}
\mlsysauthor{Stephanie Wang}{uw}
\mlsysauthor{Arvind Krishnamurthy}{uw}
\mlsysauthor{Luis Ceze}{uw}
\mlsysauthor{Baris Kasikci}{uw}
\end{mlsysauthorlist}

\mlsysaffiliation{uw}{Paul G. Allen School of Computer Science and Engineering, University of Washington, Seattle, WA, USA}
\mlsysaffiliation{sjtu}{Shanghai Jiao Tong University, Shanghai, China}
\mlsysaffiliation{harvard}{Harvard John A. Paulson School of Engineering and Applied Sciences, Harvard University, Cambridge, MA, USA}

\mlsyscorrespondingauthor{Baris Kasikci}{baris@cs.washington.edu}

\mlsyskeywords{RAG, Prefetching, Acceleration, IVF index}

\vskip 0.3in

\begin{abstract}

Retrieval-augmented generation (RAG) extends large language models (LLMs) with external data sources to enhance factual correctness and domain coverage.
Modern RAG pipelines rely on large datastores, creating a significant system challenge: achieving high throughput and low latency is difficult, especially when GPU memory is limited.
To address these challenges, we propose \sys, an efficient inference system that reduces latency and improves throughput with minimal GPU memory requirements.
The core innovation of \sys is \emph{lookahead retrieval}, a prefetching mechanism that predicts required data and transfers them from CPU to GPU in parallel with LLM generation.
In addition, \sys{} adopts a prefetching scheduler and a cache-aware scheduler to support efficient multi-GPU inference with minimal overhead.
Evaluations show \sys{} achieves up to a \cameraready{1.98$\times$} average end-to-end latency reduction (single-query) and 1.83$\times$ higher average throughput (batched), as well as good scalability in throughput.
This confirms the practical utility of \sys for faster and more memory-efficient deployments of RAG applications.

\end{abstract}

]



\printAffiliationsAndNotice{\mlsysEqualContribution \internship{}} 

\section{Introduction}
\label{sec:intro}

    

Retrieval-augmented generation (RAG) has emerged as a powerful technique to enhance large language models (LLMs) by integrating them with external databases~\cite{gao2023retrieval_rag_survey,asai2024reliable,lewis2020retrieval}.
During inference, RAG \emph{retrieves} relevant content from external data sources, usually indexed as vector datastores, to mitigate issues such as hallucinations~\cite{mallen2022not,ram2023context,khandelwal2019generalization} and incorporate up-to-date or private information~\cite{izacard2023atlas,min2023silo}.

\begin{figure}[t]
    \centering
    \includegraphics[width=0.99\linewidth]{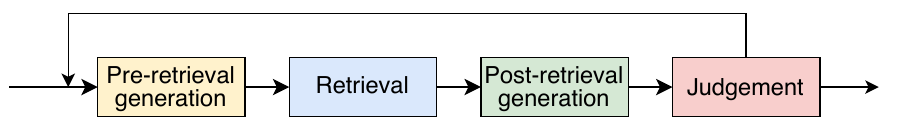}
    \vspace{-0.5em}
    \caption{Typical pipeline stages of a RAG application.} 
    \label{fig:fig1a}
    \vspace{-1.5em}
\end{figure}

To provide accurate responses, modern RAG applications employ multiple rounds of LLM calls and retrievals for a single query~\cite{gao2023retrieval_rag_survey,fan2024_rag_survey,advanced-rag-overview,modular-rag-survey,databricks-blog,azure-blog}.
As shown in Figure~\ref{fig:fig1a}, RAG pipelines typically consist of four stages:
(1)~\emph{pre-retrieval generation:} refines the initial user query to improve retrieval effectiveness.
(2)~\emph{retrieval}: finds the relevant documents from the datastore;
(3)~\emph{post-retrieval generation:} generates a response based on the user's query and retrieved documents;
(4)~\emph{judgment:} evaluates the output to determine the execution flow.


RAG datastores are typically large, as datastore size is strongly correlated with model accuracy~\cite{min2023silo,khandelwal2019generalization,borgeaud2022improving,hardt2023test,shaoscaling}.
However, a brute-force search on such a large datastore is prohibitively slow and inefficient, as most data is irrelevant to the query.
This drives the adoption of solutions like the Inverted File Index (IVF)~\cite{ivf_original}, which partitions the data into clusters and restricts the search to only the most relevant ones.


\begin{figure}[t]
    \centering
    \includegraphics[width=0.95\linewidth]{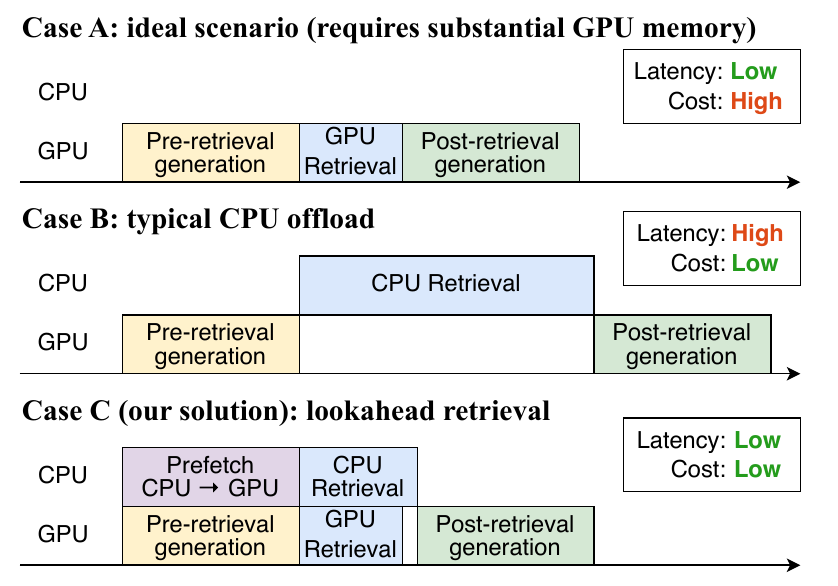}
    \vspace{-1em}
    \caption{Illustration of \sys's \textit{\lookah} mechanism and comparison to different RAG systems. \sys prefetches relevant retrieval data from CPU to GPU, overlaps data transfer with the pre-retrieval stage, and accelerates retrieval with GPU--CPU cooperation.
    }
    \label{fig:fig1b}
    \vspace{-1em}
\end{figure}

While this partitioning effectively reduces the computational burden of the search, it does not solve the memory capacity challenge.
To accelerate the IVF retrieval process with the GPU, one straightforward approach is to keep the entire datastore in the GPU memory (case A in Figure~\ref{fig:fig1b}).
As the size of a datastore is large (tens to thousands of GB)~\cite{quinn2025_iks},
such a setting is expensive and often commercially impractical for local and custom RAG deployments handling private or sensitive data.
Even in large data centers, where user requests are often batched, this approach challenges service level objectives (SLOs) because allocating GPU memory to the datastore reduces the LLM's key-value cache (KV cache), limiting effective batch sizes~\cite{kwon2023efficient}.
An alternative approach is to load the relevant data clusters from CPU to GPU at runtime.
However, because bandwidth between the CPU and GPU is typically limited, this approach suffers from high transfer latency.


As a result, modern RAG systems~\cite{jiang2025rago,shen2025hermes} process retrieval on CPUs as they typically offer large and relatively cheap memory.
However, CPUs are not as efficient as GPUs for highly parallelizable operations such as vector similarity search, and therefore using CPUs for search (case B in Figure~\ref{fig:fig1b}) increases RAG system latency.
For latency-sensitive applications such as customer chatbots~\cite{vonage_reducing_rag_latency, akkiraju2024facts, ontinue_latency_ion_iq_chatbot}, financial analysis~\cite{harchworks_rag_financial, myscale_rag_trading}, and emergency medical diagnosis~\cite{klang2024assessing, apollo_medlm_rag}, high latency can lead to poor user experience and even critical failures.
To make matters worse, any latency increase of RAG systems has compounding effects due to multiple rounds of LLM generation and retrieval.



\pgheading{Our proposal}
In this work, we observe that there is a \emph{substantial semantic overlap} between a user's initial query and the query refined by an LLM during the pre-retrieval stage (see Figure~\ref{fig:fig1a}).
Since both queries represent the same core information need, their relevant IVF clusters are also likely to overlap significantly (see \S \ref{sec:analysis-similarity} for detailed analysis).

Using this insight, we propose \sys{} (\S\ref{sec:design}), an efficient inference system that optimizes RAG performance by leveraging GPU retrieval acceleration while minimizing GPU memory consumption.
\sys{} employs \emph{\lookah} to proactively load relevant IVF clusters onto the GPU, hiding the CPU--GPU data transfer overhead during concurrent LLM generation.
As illustrated in Figure~\ref{fig:fig1b}, our approach aims to significantly reduce retrieval latency without exceeding GPU memory constraints.
To guarantee retrieval accuracy, \sys{} complements this prefetching with a hybrid search: any clusters missed by the lookahead are simultaneously searched on the CPU, and the results are merged.
In addition to prefetching, \sys{} also utilizes on-GPU caching to further reduce redundant data transfers for frequently accessed clusters.

\sys{} also optimizes for batch and multi-GPU inference (\S\ref{sec:design:batch-multi-gpu}).
To mitigate increased data transfer in batched scenarios, \sys{} uses a prefetching scheduler to group similar queries, maximizing the overlap of their prefetched clusters.
In multi-GPU settings, \sys{} employs a cache-aware query scheduler that routes queries to the appropriate GPU to maximize the utility of its cached clusters.

\pgheading{Results summary}
We evaluated \sys{} using six RAG pipelines with Llama 3B/8B models~\cite{llama3modelcard} on a Wikipedia datastore~\cite{karpukhin2020dense}.
\sys{} demonstrates significant efficiency, enabling a datastore (61\,GB) and a \llama (16\,GB) LLM to run on a single \gpusmall GPU (24\,GB), with performance far exceeding the CPU baseline.
For single-query inference on an \gpusmall, \sys{} achieves an average 1.53$\times$ end-to-end latency speedup.
Throughput gains increase with batching, reaching \cameraready{1.98$\times$} on average (batch size 8, \gpularge).
In multi-GPU settings, its prefetching and cache-aware scheduling deliver strong scaling (\eg on \gpuhuge, 3.8$\times$ speedup on 4 GPUs, compared to the performance on 1 GPU).
These results confirm that \sys{} enables scalable RAG serving under tight GPU memory constraints.
Our code is publicly available at \url{https://github.com/uw-syfi/TeleRAG}.

In summary, we make the following key contributions:
\begin{itemize}[itemsep=1pt, topsep=1pt, leftmargin=*]
\item Analysis of the correlation of queries between the pre-retrieval generation and retrieval stages, revealing significant overlap in their corresponding IVF clusters.
\item \textit{Lookahead retrieval}, which prefetches likely IVF clusters to the GPU, and hides CPU--GPU data transfer time during pre-retrieval generation.
\item \sys, an efficient RAG inference system that integrates \textit{\lookah} and supports high-performance multi-GPU inference through prefetching and cache-aware scheduling, resulting in significant acceleration of RAG with minimal GPU memory usage.

\end{itemize}

\section{Background}
\label{sec:background}

\subsection{Retrieval-Augmented Generation (RAG)}
\label{sec:background-rag}
RAG is a technique that enhances the capabilities of LLMs by integrating them with information retrieval to generate more accurate and relevant text~\cite{gao2023retrieval_rag_survey,asai2024reliable,lewis2020retrieval}. 
The core idea behind RAG is to augment the LLM with relevant information retrieved from a large corpus of documents, improving the LLM's ability to answer questions without hallucinations~\cite{mallen2022not,ram2023context,khandelwal2019generalization} and generate content based on up-to-date or private information~\cite{izacard2023atlas,min2023silo}.

\begin{figure}[t]
    \centering
    \vspace{-1em}
    \includegraphics[width=0.99\linewidth]{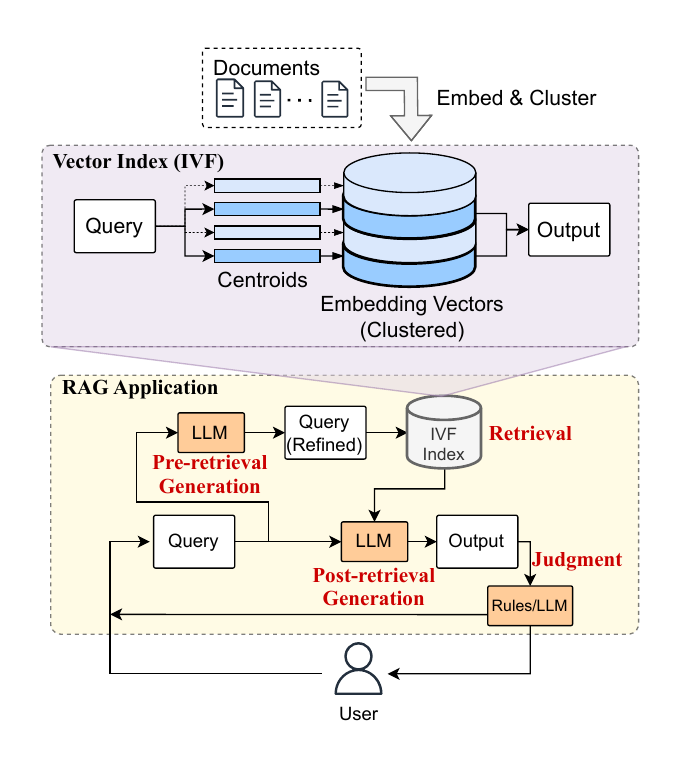}
    \vspace{-2.8em}
    \caption{Overview of RAG.}
    \label{fig:rag-overview}
    \vspace{-1.5em}
\end{figure}

\pgheading{Modern RAG pipelines} 
In order to deliver precise and contextually appropriate responses, 
modern RAG models adopt a \emph{modular} approach that employs multiple rounds of LLM calls and retrievals for a single query
~\cite{gao2023retrieval_rag_survey,fan2024_rag_survey,advanced-rag-overview,modular-rag-survey,databricks-blog,azure-blog}.
Typically, they have the following types of steps (shown in Figure \ref{fig:rag-overview}):
(1)~\textbf{Pre-retrieval generation} decides whether retrieval is needed or polishes queries before retrieval; for example, query transformation~\cite{ma2023query,jagerman2023query,gao2022precise_hyde,zhou2022least,ye2023enhancing,press2022measuring,peng2024large,zheng2023take,llamaindex} reformulates the original query to make it clearer and more suitable for retrieval. (2)~\textbf{Retrieval} then identifies relevant documents for the query, taking the output of pre-retrieval generation and producing the evidence for the next stage. (3)~\textbf{Post-retrieval generation} uses the user's query and the retrieved documents to produce the final response; it can also include processes like summarization~\cite{jiang2023longllmlingua,kim2023sure} or reranking~\cite{zhuang2023open,cohere-rerank}. (4)~\textbf{Judgment} dynamically determines the execution flow (\eg deciding whether further iteration is needed to enhance the response) using heuristics or LLMs.

\subsection{Datastore, Inverted File Index (IVF), and Retrieval}

\pgheading{Datastore}
The datastore is processed by cleaning, chunking, and indexing to enable efficient retrieval. The raw data, in various formats, is first cleaned, converted to plain text, and divided into \emph{chunks}. The chunks are then converted into vector embeddings using an embedding model such as Contriever~\cite{izacard2021unsupervised}. The document chunks, along with vector embeddings, are stored in a \emph{vector database}, which enables efficient retrieval by searching for chunks based on vector similarities between the query and the chunk embeddings.

\pgheading{IVF-based retrieval}
Many RAG deployments use the Inverted File Index (IVF) algorithm~\cite{ivf_original} for efficient vector search.
IVF partitions the vectors into \emph{clusters} based on similarity, each represented by a \emph{centroid}.
The search is then restricted to only the few clusters most relevant to the query, avoiding searching the entire datastore.

With clustering, an IVF search is a two-step process: (1)~\textbf{Cluster probing}: The query is compared only to the cluster centroids to quickly identify the few most relevant clusters.
(2)~\textbf{Searching}: The system then performs a detailed search only within that small set of selected clusters and returns the final top-$k$ best-matching items.

A parameter called \nprobe~\cite{douze2024faiss} controls the number of clusters to check in the first step. A larger \nprobe increases accuracy at the cost of higher latency, as more data must be searched. \displayappendix{B}{\ref{sec:ivf-math}} provides a formal description of IVF.

\pgheading{Latency of retrieval}
Since the search process is highly parallelizable across clusters and vectors, this search algorithm can be highly accelerated by GPUs. Furthermore, open-source libraries offer efficient GPU implementations~\cite{johnson2019billion,rapidsai}. 
However, the vector database in modern RAG deployments can reach tens to thousands of GB~\cite{quinn2025_iks, sella2024flash} as a result of the positive correlation of datastore size with accuracy~\cite{min2023silo,khandelwal2019generalization,borgeaud2022improving,hardt2023test,shaoscaling}. The large vector databases increase the latency of RAG pipelines significantly in both local setups and data centers.

Local setups typically consist of personal workstations or laptops with a single consumer-grade GPU (\eg \gpusmall has 24\,GB of memory),
where latency is critical. Data center environments, on the other hand, care about both latency and throughput since they handle numerous simultaneous user queries. In both scenarios, GPU memory capacity becomes a performance limitation when working with vector databases. Consumer GPUs in personal setups lack sufficient memory to accommodate large vector databases, whereas data-center GPUs must sacrifice KV cache space (reducing throughput) to accommodate them~\cite{kwon2023efficient}. Consequently, vector databases are commonly stored in the CPU memory instead. Unfortunately, this approach increases retrieval latency whether IVF search operations are performed on the CPU or data is transferred between CPU and GPU (as explored in \S\ref{sec:analysis}).

\section{Analyzing RAG Latency}
\label{sec:analysis}

In this section, we analyze the system challenges in state-of-the-art RAG applications.
We constructed a 61\,GB vector index with the Faiss library~\cite{douze2024faiss}, and set the IVF clusters to 4096 following common practice~\cite{asai2023self}. 
Our analysis is conducted on an \gpusmall (24\,GB) and \gpularge (80\,GB) using \llama~\cite{llama3modelcard}.
Further setup details are in \S\ref{sec:exp_setups}.

\subsection{End-to-end Latency of RAG Pipelines}
\label{sec:analysis-e2e-latency}

\begin{figure}[t]
    \begin{subfigure}[t]{0.47\linewidth}
        \centering
        \includegraphics[width=\linewidth]{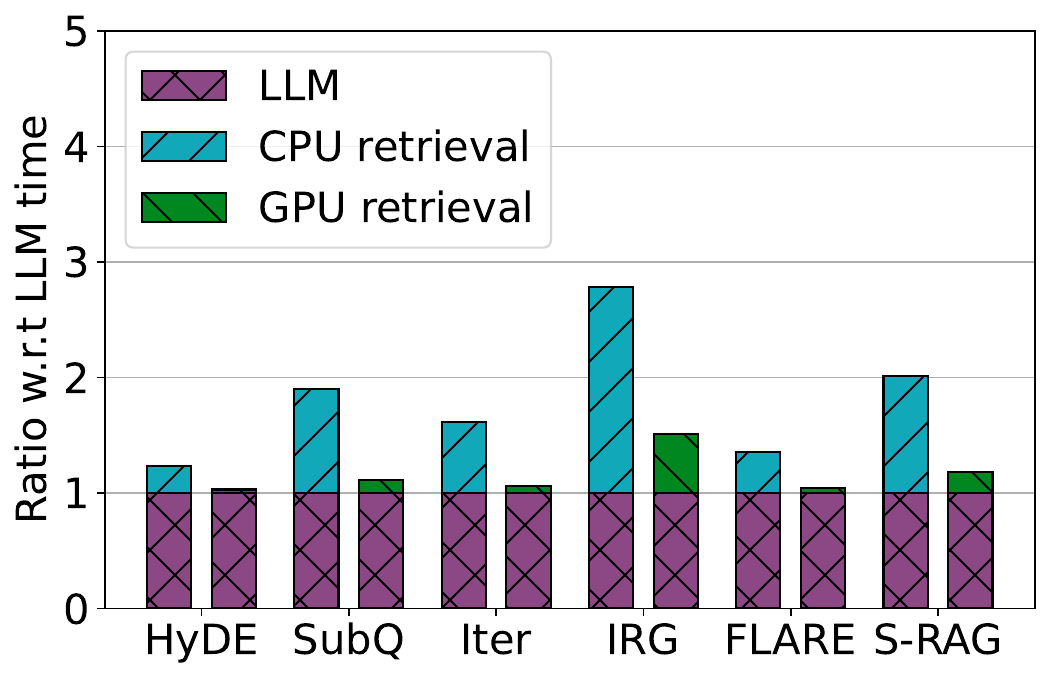}
        \vspace{-1.5em}
        \caption{Batch = 1.}
    \end{subfigure}
    \begin{subfigure}[t]{0.45\linewidth}
        \centering
        \includegraphics[width=\linewidth]{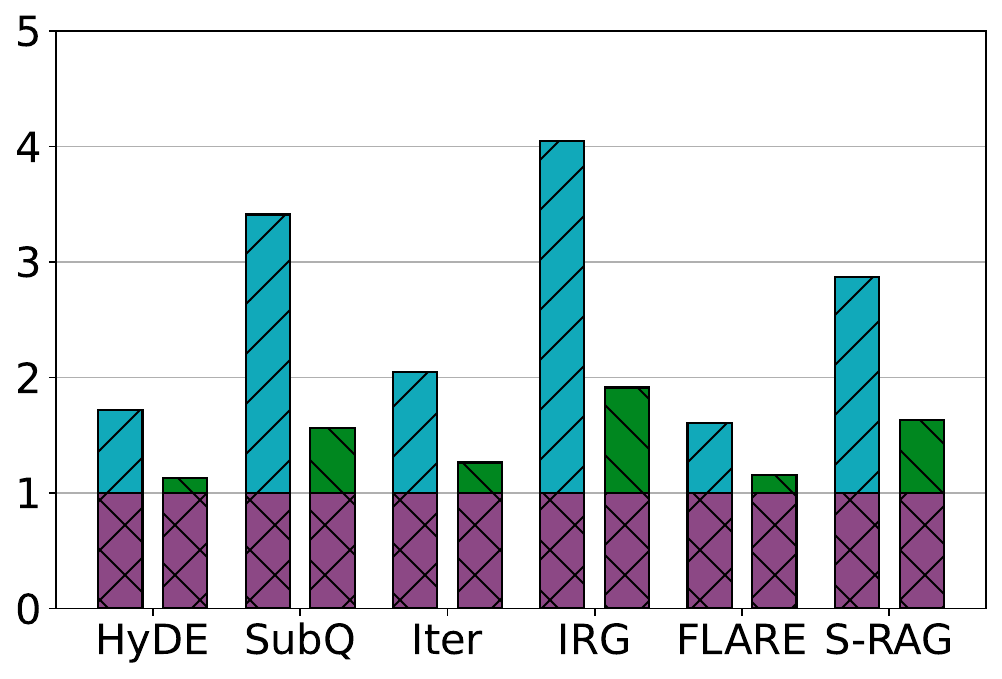}
        \vspace{-1.5em}
        \caption{Batch = 4.}
    \end{subfigure}
    \vspace{-0.7em}
    \caption{Latency breakdown of six RAG pipelines on NQ dataset~\cite{kwiatkowski2019natural} with one \gpularge GPU.}
    \label{fig:rag-breakdown-time}
    \vspace{-1.5em}
\end{figure}

We first analyze the end-to-end latency of six RAG pipelines (\S\ref{sec:dataset_setups}) by comparing two scenarios: (1)~LLM on GPU with retrieval on CPU (low GPU memory), and (2)~both LLM and the vector index on GPU (low latency).

We set the \nprobe to 256, a commonly used setting under this index's configuration (see \cameraready{\S\ref{sec:dataset_setups}} and \S\ref{sec:exp_setups} for details).
As shown in Figure~\ref{fig:rag-breakdown-time} (using randomly sampled 1024 NQ queries~\cite{kwiatkowski2019natural}), the CPU-based retrieval phase is a major bottleneck, consuming 41.1\% and 60.5\% of total latency for batch sizes 1 and 4, respectively.
In contrast, GPU-accelerated retrieval accounts for only 10.5\% and 28.3\% of the latency, respectively.
On average, GPU retrieval is 5.96$\times$ (batch size 1) and 3.87$\times$ (batch size 4) faster than its CPU counterpart, reducing overall end-to-end latency by 1.5$\times$ and 1.8$\times$.
Thus, accelerating GPU-based retrieval is crucial for reducing end-to-end latency.

However, GPU acceleration has a significant memory cost.
For instance, in our setting, with a 61\,GB index and a 16\,GB \llama model, GPU retrieval requires 77\,GB of GPU memory, far exceeding the 24\,GB of a common \gpusmall.
This makes GPU-accelerated retrieval often unfeasible on lower-end GPUs or with large datastores.

Even on large GPUs (\eg \gpularge), storing the full index in memory limits serving throughput.
High-throughput batched inference is bottlenecked by the LLM's KV cache capacity \cite{kwon2023efficient}.
The index's large memory footprint reduces the available memory for this KV cache, thereby limiting the effective batch size.
This memory contention is exacerbated by RAG's typically long contexts \cite{jin2024ragcache,lu2024turborag,yao2024cacheblend} and the growing size of models and indexes \cite{min2023silo,khandelwal2019generalization,borgeaud2022improving,hardt2023test,quinn2025_iks}, necessitating offloading.

Therefore, in the rest of this section, we try to answer the following question: \emph{Is it possible to achieve the latency of GPU-based retrieval while using much less GPU memory?}

\begin{figure}[t]
    \centering
    \includegraphics[width=0.75\linewidth]{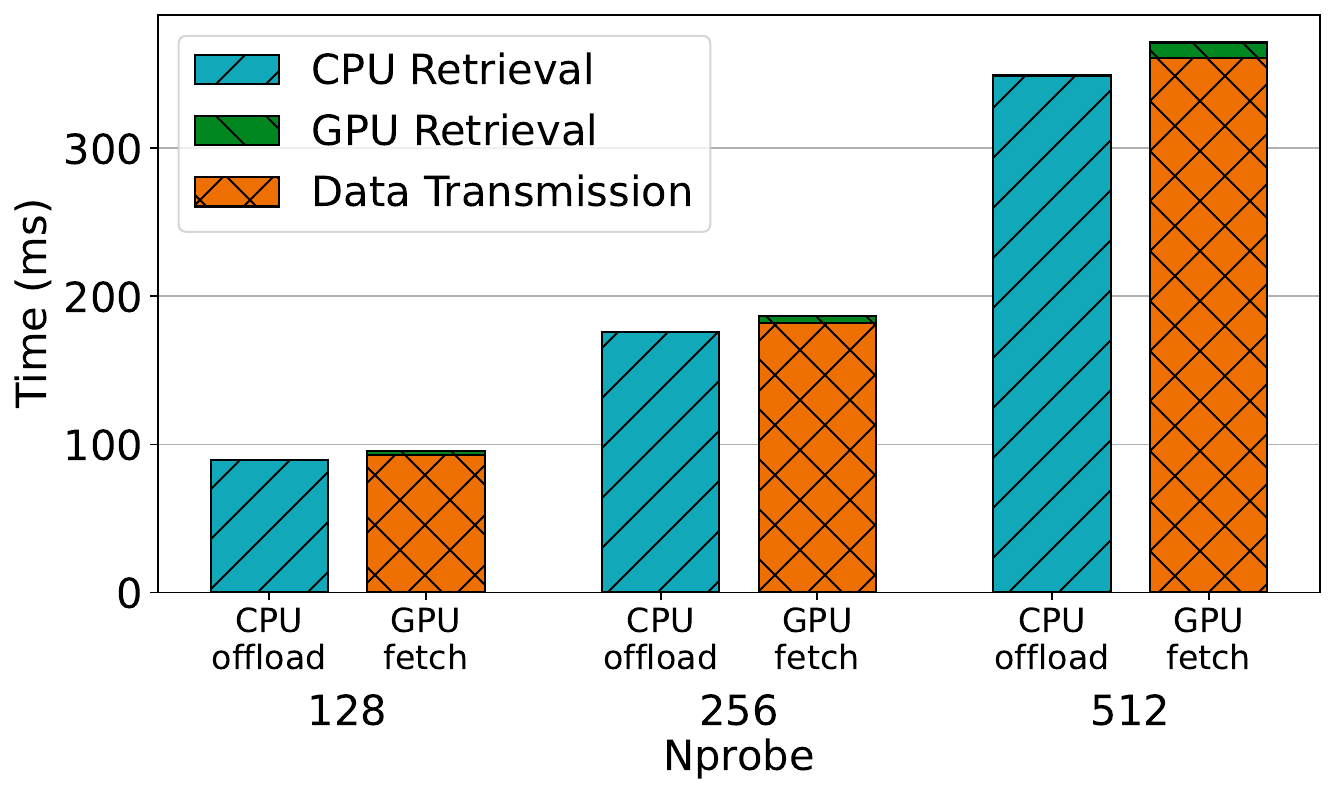}
    \vspace{-1.2em}
    \caption{Latency breakdown of CPU-offload and runtime-fetch GPU retrieval, averaged over 512 random NQ queries.}
    \label{fig:gpu-runtime-overhead}
    \vspace{-.7em}
\end{figure}

\subsection{GPU-accelerated Retrieval with Runtime Transfer}
\label{sec:analysis-retrieval}

A straightforward approach to enable GPU retrieval with limited GPU memory is to fetch the necessary data from CPU to GPU on demand at runtime, leveraging the IVF index to narrow the search space.
While this enables fast GPU search, the data transfer itself becomes the new bottleneck.

Figure~\ref{fig:gpu-runtime-overhead} compares the latency of the runtime fetching system against CPU retrieval on an \gpusmall GPU with three different \nprobe values that determine the amount of data fetched.
Overall, fetch time dominates latency due to the limited CPU--GPU PCIe bandwidth (32\,GB/s). 
Although the GPU search is substantially faster, the fetch overhead results in a higher end-to-end latency ($\sim${}3\% slower on average).
Thus, to achieve a meaningful speedup with this approach, the data-fetching latency must be effectively hidden.

To hide data-fetch costs, CPU-to-GPU transfers must occur before the retrieval stage, which requires predicting which data will be accessed.
Fortunately, modern RAG pipelines offer a valuable hint: the query from the previous step.

\begin{table}[t]

\resizebox{\columnwidth}{!}{
    \centering
    \begin{tabular}{l|cccccc}
        \toprule
        Dataset & HyDE & SubQ & Iter & IRG & FLARE & S-RAG \\
        \midrule
        NQ & 73.1\% & 63.2\% & 91.5\% & 83.8\%   & 79.1\%  & 100.0\% \\
        HotpotQA & 75.3\% & 62.5\% & 89.6\% & 89.4\%   & 80.2\%  & 100.0\% \\
        TriviaQA & 73.1\% & 61.6\% & 86.2\% & 86.1\%   & 81.7\%  & 100.0\% \\ 
        \bottomrule
        \end{tabular}
    }
    \vspace{-0.7em}
    \caption{IVF cluster overlapping rate between the input and output of the pre-retrieval generation, \cameraready{\ie the average percentage of correctly predicted clusters among the clusters actually used.} Since Self-RAG does not incorporate query transform, its coverage is always $100\%$.}
    \label{tab:analysis-coverage}
    \vspace{-2em}
\end{table}

\begin{figure*}[t!]
    \centering
    \includegraphics[width=.98\linewidth]{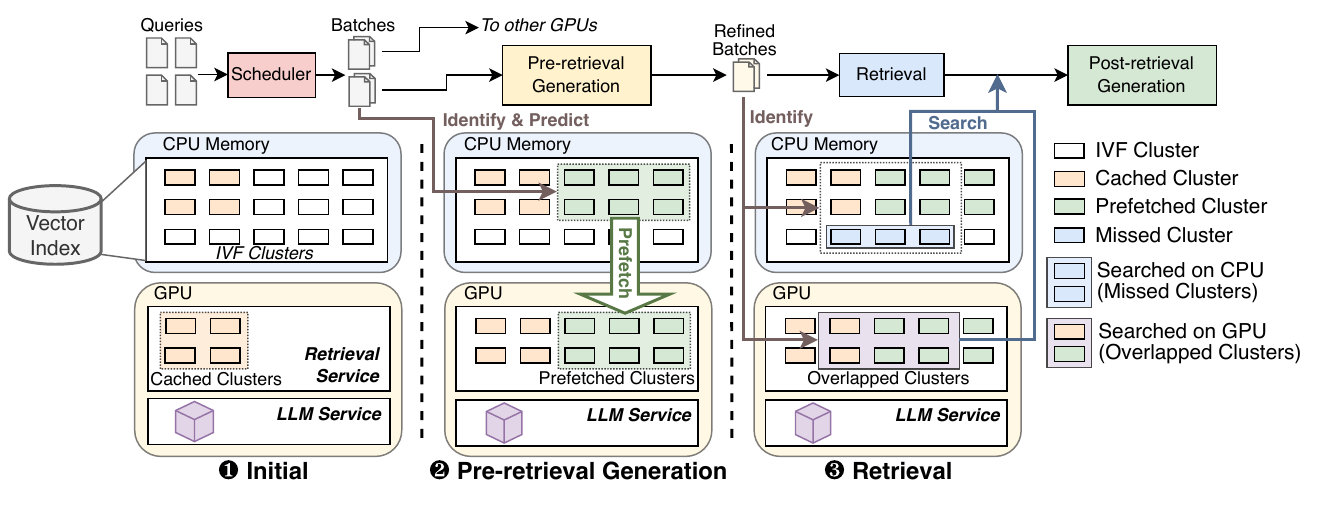}
    \vspace{-1.3em}
    \caption{The overview of \sys{} system.
    When queries arrive, some clusters are already cached on the GPU from previous requests.
    These queries are grouped into micro-batches based on semantic similarity.
    For each micro-batch, \sys{} prefetches the required clusters using \emph{\lookah{}} in parallel with the pre-retrieval generation.
    During retrieval, \sys{} performs a hybrid search: the GPU processes cache hits directly from GPU memory, while the CPU handles cache misses.
    The retrieved documents are then passed to the post-retrieval generation stage.
    }
    \label{fig:full_system_overview}
    \vspace{-0em}
\end{figure*}

\subsection{Overlapping of IVF Clusters}
\label{sec:analysis-similarity}
While the exact data to be retrieved is only known after the pre-retrieval generation is done, we observe high similarity in the IVF cluster assignments between queries at different stages.

\pgheading{Similarity of queries at different stages}
During RAG pre-retrieval stages (\eg query transformation~\cite{gao2022precise_hyde, zheng2023take, peng2024large, openscholar}),
an LLM refines an initial user query \qin into a transformed query \qout for the actual retrieval.
While this process often rewrites or simplifies the query, it intuitively preserves its core semantic content.
Consequently, the embedding vectors of \qin and \qout are likely to be similar,
suggesting their target IVF clusters will significantly overlap.
Therefore, \qin can serve as a valuable hint for predicting \qout.

\pgheading{Prediction coverage} 
To verify this hypothesis, we evaluated the average cluster coverage (predicted vs.\ actual) in three popular QA datasets (NQ~\cite{kwiatkowski2019natural}, HotpotQA~\cite{yang2018hotpotqa}, and TriviaQA~\cite{joshi2017triviaqa}) and six RAG pipelines. As Table~\ref{tab:analysis-coverage} shows, when prefetching 256 clusters, the overlap is consistently high. For instance, even the lowest reported coverage (for SubQ) remains above 61.6\%.

\pgheading{Opportunity}
This data shows an opportunity to predict required clusters, hiding data transfer overhead during LLM generation.
In this paper, we aim to leverage this observation to accelerate the inference latency for RAG.

\section{Design of \sys}
\label{sec:design}

Driven by the need for high-performance but memory-efficient RAG inference, we propose \sys{}, an end-to-end system that accelerates retrieval through its core \emph{\lookah{} mechanism}, which accelerates the retrieval process with a GPU but only requires a small fraction of the datastore on the GPU memory.
Alongside this technique, \sys{} further designs optimizations on prefetching amount, caching, and scheduling algorithms for batching and multi-GPU, forming a complete system for both local and cloud use cases.

This section details the \lookah{} mechanism (\S\ref{sec:design:lookahead_retrieval}), and then describes how \sys{} extends it with batching and multi-GPU support (\S\ref{sec:design:batch-multi-gpu}), followed by other system-level optimizations (\S\ref{sec:design:system_optimizations}). \displayappendix{D}{\ref{sec:impl}} provides implementation details.

\subsection{Lookahead Retrieval Mechanism}
\label{sec:design:lookahead_retrieval}

The core mechanism of \sys{}, \lookah{}, is inspired by the observation in \S\ref{sec:analysis-similarity} that queries across different RAG stages are highly correlated and therefore select overlapping IVF clusters. 
Building on this insight, \lookah{} predicts and prefetches likely-needed IVF clusters to the GPU during LLM generation, overlapping data transfer with computation. 
During retrieval, \sys{} leverages \lookah{} to coordinate the CPU and GPU so that prefetched data are processed on the GPU, while the CPU concurrently handles any missed clusters. 
This cooperative execution forms the foundation of \sys{}’s low-latency, memory-efficient design.

Figure~\ref{fig:full_system_overview} shows the overview of \sys{} with \lookah{}. 
Let \qin denote the query input to the pre-retrieval stage and \qout denote the output passed to the retrieval stage. 
The IVF clusters selected by \qin and \qout are highlighted with a green background (\Cin) and a purple background (\Cout), respectively. 
Due to the semantic similarity between \qin and \qout, there is substantial overlap between \Cin and \Cout.

Guided by this correlation, \lookah{} operates in the following steps: 
\begin{enumerate}[itemsep=0pt, topsep=1pt, leftmargin=*]
    \item \textbf{Predict \& prefetch:} During LLM generation, prefetch IVF clusters likely to be used in retrieval to GPU memory, identified by their distance to \qin. This transfer is performed asynchronously via GPU DMA, overlapping with ongoing LLM computation.
    \item \textbf{GPU similarity search:} Once \qout is available, the GPU efficiently searches the predicted clusters (\Coverlap) already resident in GPU memory.
    \item \textbf{CPU similarity search:} Concurrently, the CPU performs similarity search over the remaining clusters (\Cmiss) that were not prefetched.
    \item \textbf{Merge:} The search results from GPU and CPU are merged on GPU. The retrieval documents are fed to LLM for post-retrieval generation on GPU.
\end{enumerate}

In summary, \lookah{} enables \sys{} to accelerate retrieval by overlapping data prefetching with LLM generation and distributing the similarity search between GPU and CPU. 
This design significantly reduces CPU computation and data transfer latency, forming the backbone of \sys{}’s efficiency.

\pgheading{Prefetching amount}
\label{sec:design:optimal-amount}
Prefetching more clusters improves hit rate and retrieval speed, but also increases CPU--GPU transfer time. If this transfer exceeds the pre-retrieval generation window, the overlap advantage is lost and may add latency, though slight overruns are acceptable if retrieval savings are substantial.
Through analytical modeling and empirical profiling on modern hardware, we found \emph{prefetching up to the pre-retrieval LLM generation time} ($t_{\mathrm{LLM}}$) achieves the optimal balance between latency reduction and transfer overhead. 
A detailed derivation of this result is provided in \displayappendix{C}{\ref{sec:math-analysis-prefetch-amount}}.
Since $t_{\mathrm{LLM}}$ varies per query, for a given pipeline and hardware bandwidth $B_{\mathrm{link}}$, we estimate its average ($\overline{t}_{\mathrm{LLM}}$) on a calibration set and set the prefetching amount to $B_{\mathrm{link}} \times \overline{t}_{\mathrm{LLM}}$.

\subsection{Batching and Multi-GPU System}
\label{sec:design:batch-multi-gpu}

\begin{figure}[t!]
    \centering
    \includegraphics[width=0.98\linewidth]{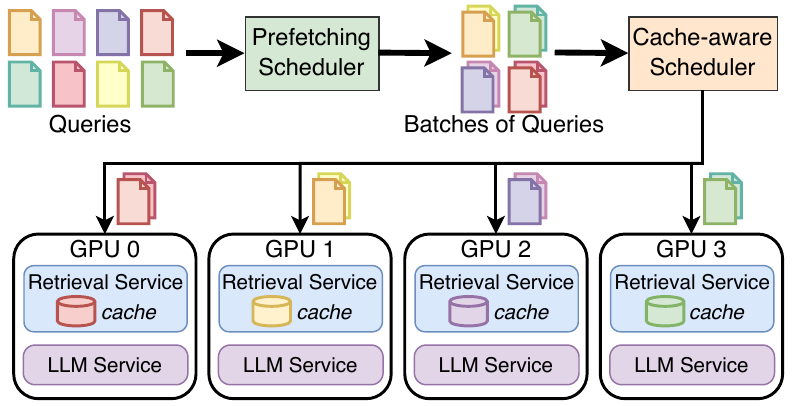}
    \vspace{-0.6em}
    \caption{Overview of \sys{}’s multi-GPU system design. The prefetching scheduler clusters semantically similar queries into micro-batches, while the cache-aware scheduler assigns these batches to GPUs based on cache locality.}
    \label{fig:multi-gpu-design}
    \vspace{-1em}
\end{figure}

While \S\ref{sec:design:lookahead_retrieval} describes \lookah on a single-query and single-GPU basis,
we now demonstrate how \sys{} extends this mechanism to support batches and multi-GPU inference.
The system's multi-GPU architecture is illustrated in Figure~\ref{fig:multi-gpu-design}.

\pgheading{Batching design}
For batched scenarios, a challenge for \lookah is that each query requires a different set of IVF clusters.
In our design, we apply the fixed total prefetching budget as described in \S\ref{sec:design:optimal-amount}, and distribute it equally among all queries in the batch.
Although the per-query hit rate drops as the batch size grows, this strategy provides balanced retrieval acceleration for each query.

\pgheading{Multi-GPU support}
In multi-GPU scenarios that serve a large number of RAG queries,
\sys{} divides a global batch of queries into micro-batches and assigns them to each available GPU.
Each GPU then independently processes the full RAG pipeline for its assigned micro-batches.

\pgheading{Scheduling optimization for prefetching}
To compensate for the reduced hit rate in larger batches, we introduce a \textit{prefetching scheduler} (see Figure~\ref{fig:multi-gpu-design}).
It operates by performing a greedy search on the global batch, grouping queries by the lowest L2 distance.
While this scheduling adds latency, the greedy search is highly efficient on GPUs, incurring negligible overhead.
Our profiling shows that for batch sizes up to 256, the search latency is less than 0.1\,s, which is minimal compared to the end-to-end batch latency.

\pgheading{Cache design}
To improve performance, \sys{} implements a dynamic caching strategy for data clusters.
Instead of discarding all clusters after a request is served, it retains the most frequently accessed ones in GPU memory.
This allows subsequent requests to reuse these cached clusters, boosting the cache hit rate and prefetching efficiency.

\pgheading{Cache-aware scheduling}
To further exploit the benefits of caching, we design a cache-aware scheduler (Figure~\ref{fig:multi-gpu-design}) that assigns micro-batches to GPUs based on cache locality.
The scheduler employs a greedy strategy: it first selects the micro-batch with the greatest overlap across all GPU caches and assigns it to the best-matching GPU.
It then iteratively schedules the remaining micro-batches in descending order of overlap with each GPU's cached clusters.
Although this scheduling introduces minor overhead, the performance gains from improved cache locality outweigh the cost (\S\ref{sec:evaluation-analysis}).

\subsection{System Optimizations}
\label{sec:design:system_optimizations}

\pgheading{GPU sorting}
\sys accelerates the final sorting stage of IVF search by leveraging the GPU.
To achieve this, \sys transfers the scalar distance values of \Cmiss from the CPU to the GPU.
Unlike transferring full vector data, this operation is lightweight and incurs negligible overhead.
The GPU then performs a global sort over the combined distances of \Cmiss and \Coverlap.

\pgheading{Prefetching target}
\sys{} prefetches a fixed byte budget ($b_p$), rather than a cluster count, to ensure predictable transfer times given highly uneven cluster sizes.
The system fills this budget by adding whole clusters sequentially based on query proximity.
If the next closest cluster exceeds the remaining budget, it is skipped entirely, ensuring a clean GPU/CPU processing split.

\pgheading{Prefetching for multi-round}
For multi-round RAG involving the same input query, the system performs a full prefetch (up to the budget) only in the first round, leveraging high cluster similarity across rounds. In subsequent rounds, it incrementally fetches only the additional required clusters that were not loaded previously, optimizing data transfer while respecting the memory budget.

\section{Evaluation}
\label{sec:eval}
We conducted extensive experiments to evaluate the effectiveness of \sys.
In this section, we describe the necessary details on evaluation setups, present experimental results, and provide in-depth analysis and discussions.

\subsection{Evaluation Datasets and RAG Models}
\label{sec:dataset_setups}
\indent 
\pgheading{Datastore}
We built a datastore based on the wiki\_dpr dataset~\cite{karpukhin2020dense}, a popular dataset that contains 2.1 billion tokens from Wikipedia.
Following previous works~\cite{asai2023self,karpukhin2020dense,min2023silo}, we chunked the passages by every 100 tokens, and used Contriever~\cite{izacard2021unsupervised} to generate an embedding for each chunk. The embeddings have a hidden dimension of 768.
The index size is 61\,GB and we clustered the embeddings into 4096 IVF clusters.
See \displayappendix{E}{\ref{sec:detailed-index-config}} for detailed configurations of the index.

\begin{figure}[tb]
    \centering
    \includegraphics[width=.95\columnwidth]{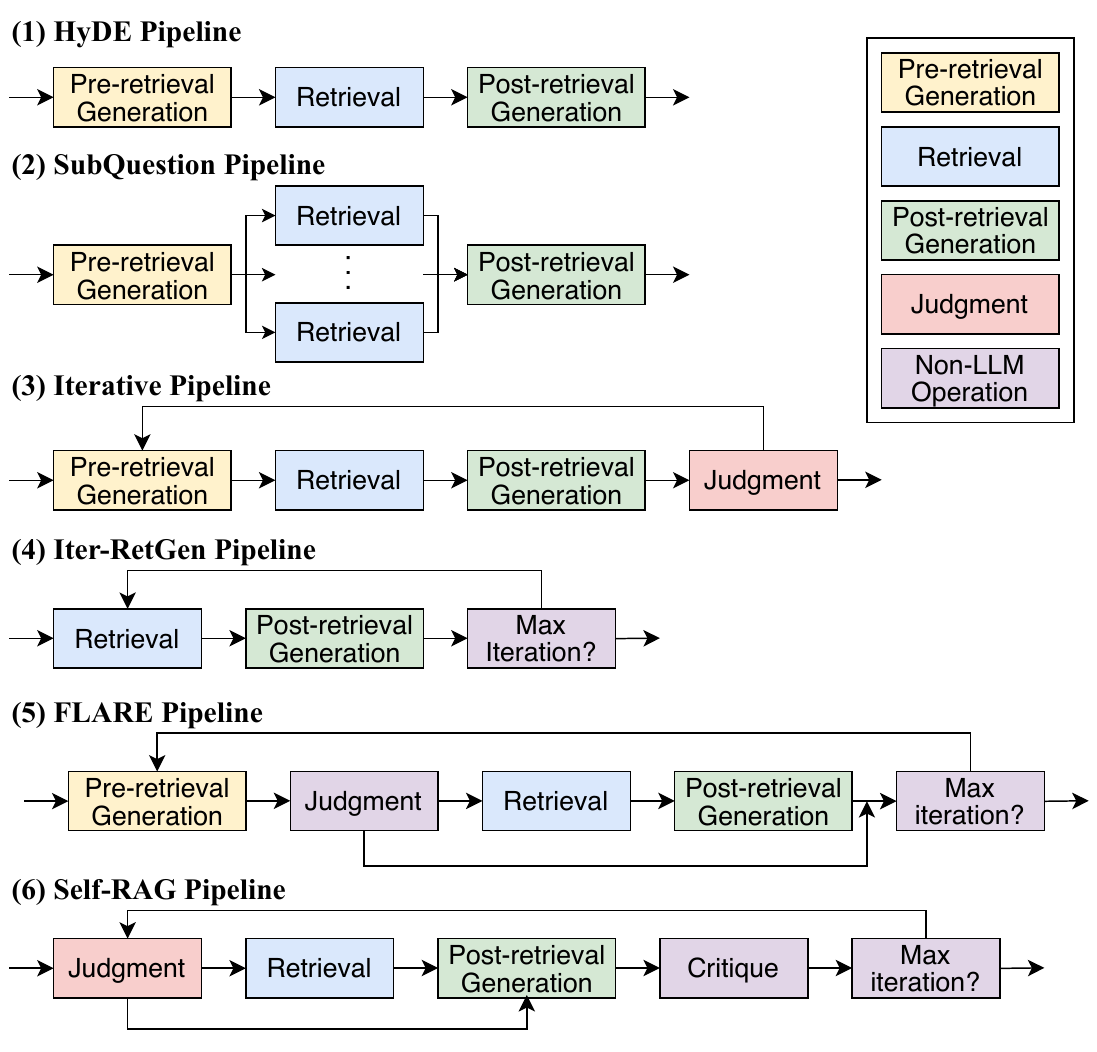}
    \vspace{-1.1em}
    \caption{Overview of \npipeline RAG pipelines that we evaluate.}
    \label{fig:workload}
    \vspace{-1em}
\end{figure}

\pgheading{LLMs}
We evaluated \sys on \llamasmall, \llama \cameraready{and \mistral} to represent different use cases.

\pgheading{RAG pipelines}
We evaluated \sys{} with \npipeline{} popular RAG pipelines, as depicted in Figure~\ref{fig:workload}.
Note that even in pipelines lacking a pre-retrieval stage, the post-retrieval generation serves a similar function for the next retrieval iteration.
Below are brief descriptions of these pipelines.
\begin{enumerate}[itemsep=0pt, topsep=1pt, leftmargin=*]
    \item \textbf{HyDE}~\cite{gao2022precise_hyde} prompts LLM to generate a hypothetical paragraph and perform retrieval based on the embedding of the generated paragraph.
    \item \textbf{SubQuestion (SubQ)}
    ~\cite{sub_question_llamaindex} 
    prompts LLM to generate multiple sub-questions and performs retrievals for each generated sub-question.
    \item \textbf{Iterative (Iter)}
    ~\cite{multistep_llamaindex} 
    prompts LLM to generate narrower questions first and iteratively refine them based on previous answers. At the end of each iteration, it prompts LLM to judge if the answer is good enough.
    \item \textbf{Iter-RetGen (IRG)} \cite{shao2023enhancing} iteratively does retrieval and LLM generation for 3 iterations. 
    \item \textbf{FLARE}~\cite{jiang2023flare} iteratively issues retrievals based on the confidence (probability score) of predicted tokens for the upcoming sentence.
    \item \textbf{Self-RAG (S-RAG)}~\cite{asai2023self} uses the LLM to judge for retrieval, generate responses, and self-critique on the responses. 
    We use the fine-tuned model based on Llama-2-7B from their official repository~\cite{selfrag_code} for trace generation.
\end{enumerate}

\pgheading{Evaluation datasets}
We use three commonly used question-answering datasets, NQ~\cite{kwiatkowski2019natural}, HotpotQA~\cite{yang2018hotpotqa}, and TriviaQA~\cite{joshi2017triviaqa}.
For each dataset, we randomly sampled 1024 queries and reported the average. When cache is enabled, to ensure stable results and avoid an initial low hit rate from a cold start, we use 512 queries for warming up the cache and another different 512 queries for evaluation.

\begin{table}[tb]
    \centering
    \resizebox{0.98\columnwidth}{!}{
        \begin{tabular}{c|ccc}
            \toprule
            Setup & \texttt{Desktop}\quad & \texttt{Server1} & \texttt{Server2} \\
            \midrule
            CPU & Threadripper\,5975 & EPYC\,9554 & EPYC\,9534 \\
            CPU mem. size & 512\,GB & 1.5\,TB & 1.5\,TB \\
            \midrule
            GPU & \gpusmall & \gpularge & $8 \times $\gpuhuge \\
            GPU mem. size & 24\,GB & 80\,GB & 140\,GB \\
            \midrule
            CPU--GPU Bus & PCIe\,4 & PCIe\,5 & PCIe\,5 \\
            Bandwidth & 32\,GB/s & 64\,GB/s & 64\,GB/s \\
            \bottomrule
        \end{tabular}
    }
    \vspace{-0.5em}
    \caption{Hardware specifications for our setups.}
    \vspace{-2em}
\label{tab:hardware-setup}
\end{table}

\subsection{Experiment Setups}
\label{sec:exp_setups}
\begin{sloppypar} 
\indent
\pgheading{Hardware setups}
We evaluated \sys on three hardware environments, \desktop{}, \serverone{} and \servertwo{}, which are equipped to represent the settings for the desktop and data center use cases.
The \desktop has the \gpusmall, and we test with 3B and 8B models.
\serverone{} and \servertwo{} are equipped with \gpularge and \gpuhuge GPUs, and we evaluate with 8B and 22B models.
We use \serverone for single GPU evaluation and \servertwo (8 GPUs) for multi-GPU evaluation.
Table~\ref{tab:hardware-setup} summarizes the hardware configurations.
\end{sloppypar}

\pgheading{Nprobe and top-$k$}
A common heuristic for the IVF index is to set \nprobe{} to $4\sqrt{N_c}$~\cite{zilliz-blog}. 
Given our index size of $N_c = 4096$, we use \nprobe{} = 256 ($ = 4\sqrt{4096}$) by default, unless otherwise specified. 
For retrieval, we use top-$k$ = 3 for the number of documents to return, unless otherwise specified.

\pgheading{RAG pipeline implementation}
We implemented the RAG pipelines with the FlashRAG framework~\cite{jin2024flashrag}. 
For IRG, FLARE, and S-RAG, we used the framework's default implementations. For the other pipelines, we reimplemented them using FlashRAG's APIs.

\pgheading{Benchmark methodology}
We follow the benchmarking methodology of SGLang~\cite{zheng2024_sglang}. Specifically, we use GPT-3.5-Turbo~\cite{gpt3-5-turbo} to execute each pipeline once and record the input and output text for every step.

\camerareadytwo{
During latency evaluation, we run real LLM inference with full KV cache allocated and perform actual autoregressive decoding, but stop generation once the number of output tokens matches the recorded trace. This lets us measure real inference performance while keeping the decoding workload fixed, ensuring a fair latency comparison across different LLM models.
}

\pgheading{Baseline systems}
To evaluate the latency of each pipeline, we constructed a clean execution flow in Python that only contains LLM generation, datastore retrieval, and other necessary logical operations to fulfill each pipeline.
For LLM generation, we used SGLang~\cite{zheng2024_sglang}, which is a state-of-the-art LLM inference engine. For retrieval, we used the industry standard retrieval library, Faiss~\cite{douze2024faiss}, as the CPU-offloaded baseline.


\pgheading{Prefetching budget setups}
Based on the methodology we described in \S\ref{sec:design:optimal-amount}, we profiled each RAG pipeline with 64 random samples from NQ~\cite{kwiatkowski2019natural} and derived the prefetching budget of each pipeline.
Since \gpularge and \gpuhuge have the same CPU--GPU bandwidth, we set the same prefetch budget for \serverone{} and \servertwo{}.

\begin{figure}[tb]
    \centering
    \begin{subfigure}[t]{0.9\columnwidth}
        \centering
        \includegraphics[width=0.95\linewidth]{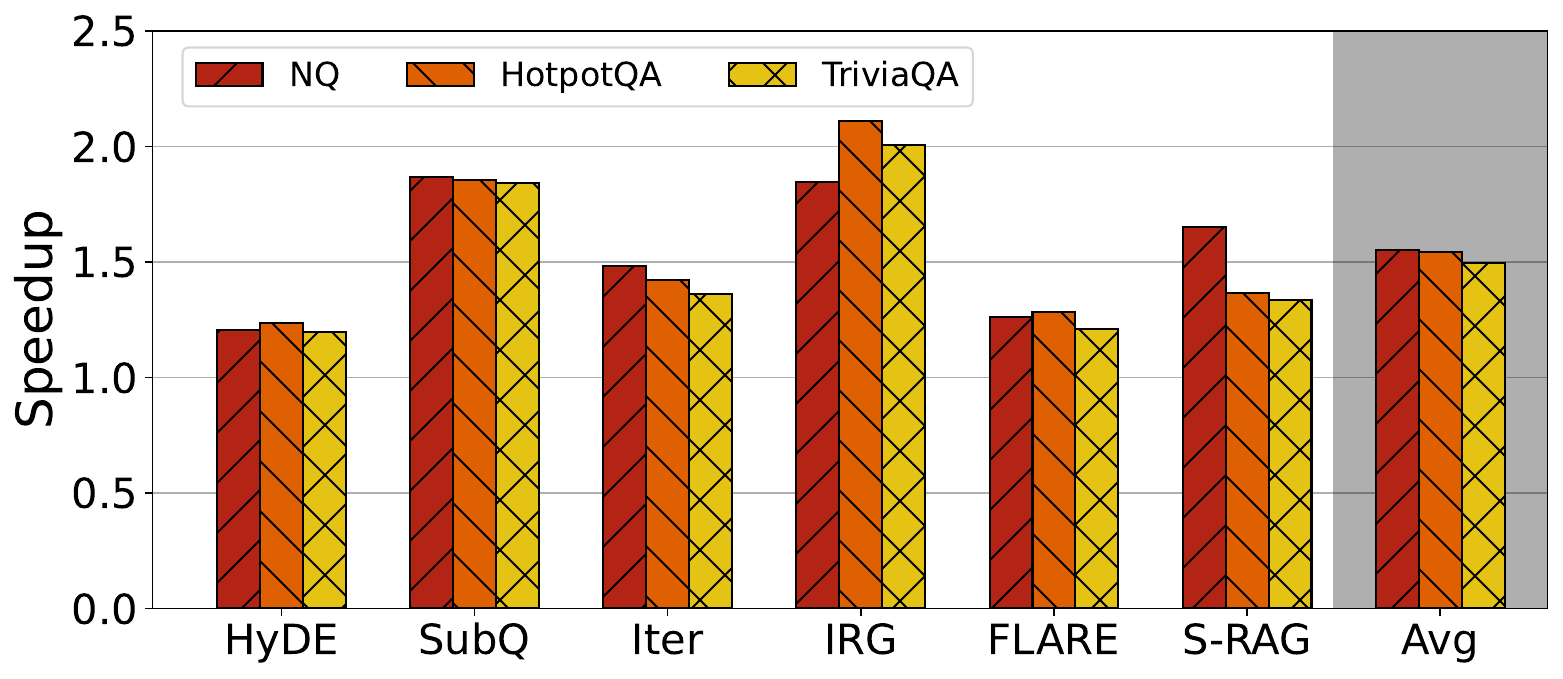}
        \vspace{-0.5em}
        \caption{End-to-end latency speedup with \llamasmall.}
    \end{subfigure}
    \begin{subfigure}[t]{0.9\columnwidth}
        \centering
        \includegraphics[width=0.95\linewidth]{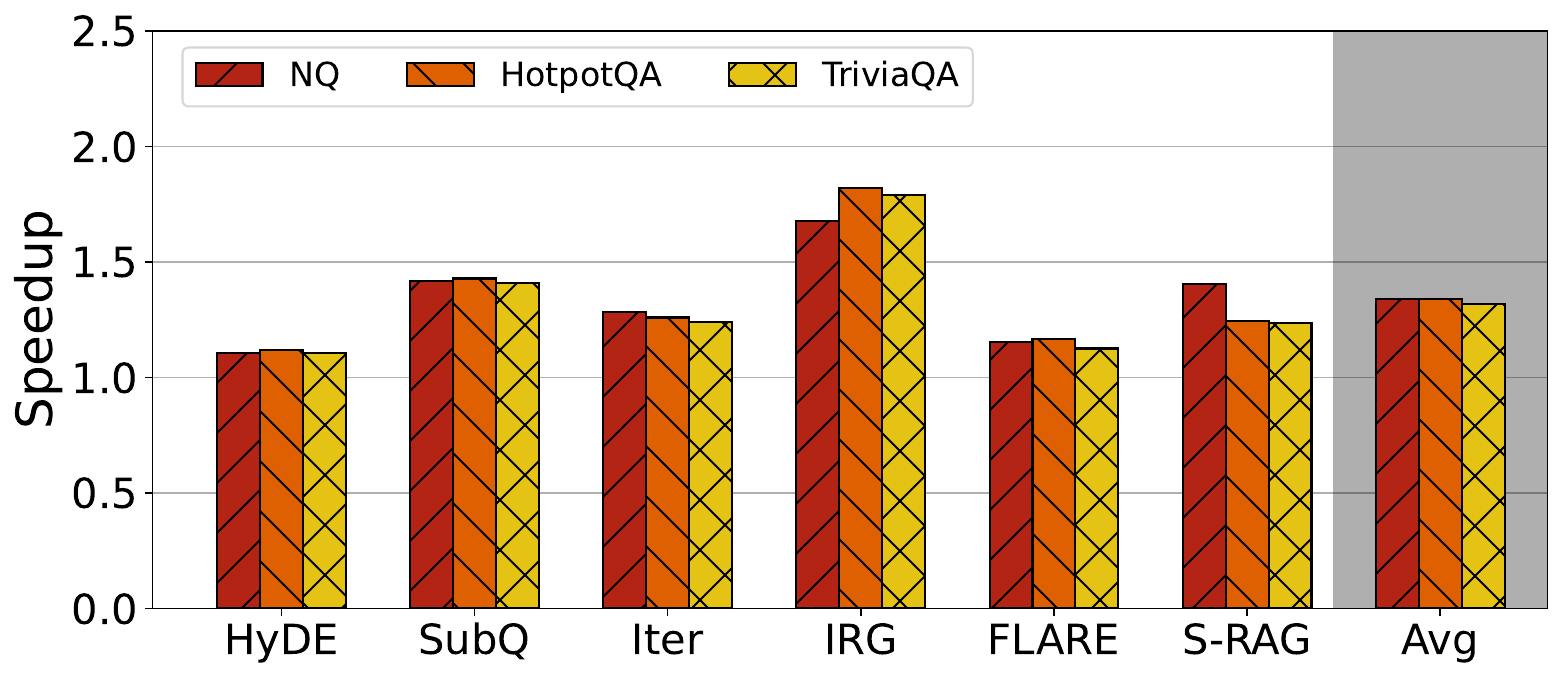}
        \vspace{-0.5em}
        \caption{End-to-end latency speedup with \llama.}
    \end{subfigure}
    \vspace{-5pt}
    \caption{End-to-end latency speedup of \sys and baseline on six RAG pipelines and three datasets, with an \gpusmall GPU.}
    \label{fig:result-speedup-rtx4090}
    \vspace{-1.5em}
\end{figure}

\begin{figure*}[t!]
    \centering
    \begin{subfigure}[t]{0.98\linewidth}
    \includegraphics[width=\linewidth]{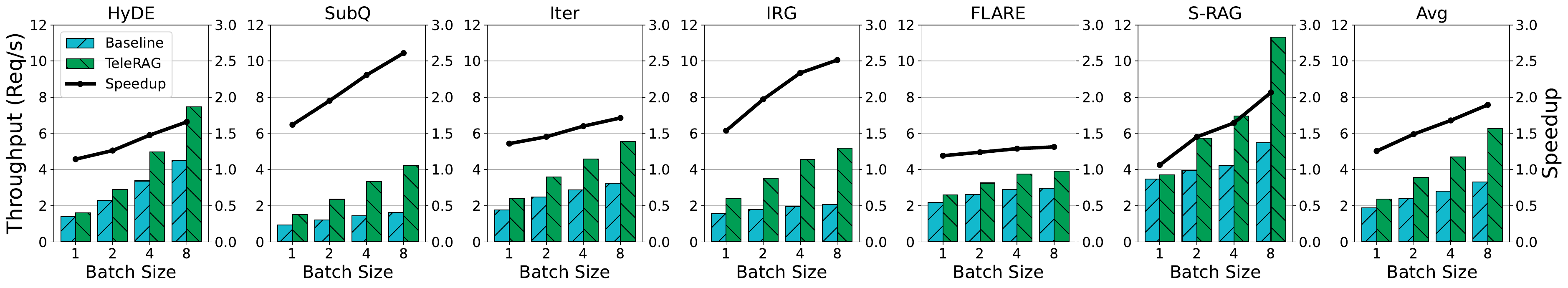}
    \vspace{-1.6em}
    \caption{End-to-end throughput on \llama.}
    \end{subfigure}
    \begin{subfigure}[t]{0.98\linewidth}
    \includegraphics[width=\linewidth]{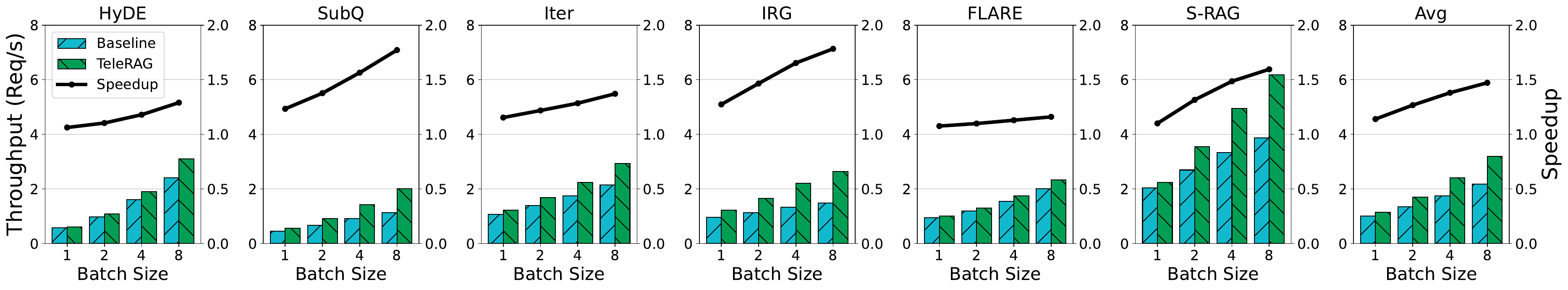}
    \vspace{-1.6em}
    \caption{\cameraready{End-to-end throughput on \mistral.}}
    \end{subfigure}
    \vspace{-0.5em}
    \caption{End-to-end throughput of \sys and the CPU-offload baseline across six RAG pipelines on the NQ dataset, evaluated at different batch sizes on an \gpularge GPU.}
    \label{fig:result-throughput-h100}
    \vspace{-0.1em}
\end{figure*}

\begin{figure*}[tb]
    \centering
    \begin{subfigure}[t]{0.32\textwidth}
        \centering
        \includegraphics[width=\linewidth]{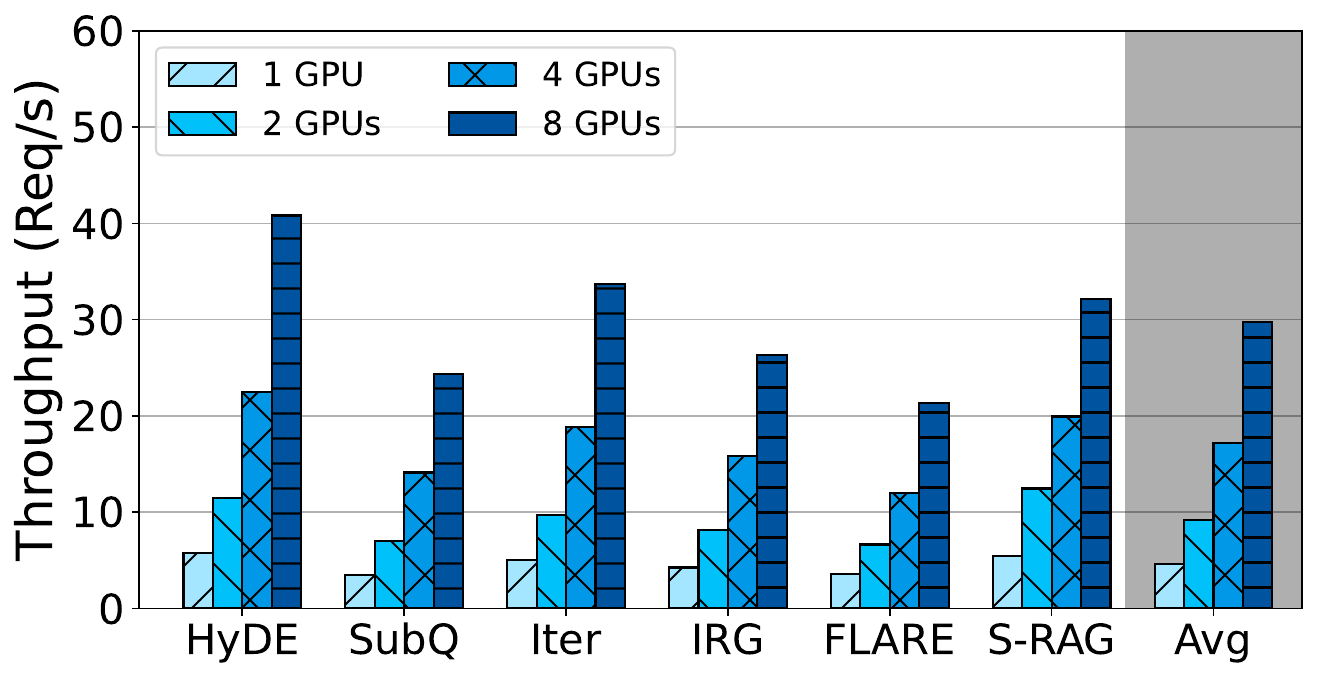}
        \vspace{-1.3em}
        \caption{End-to-end throughput on NQ.}
    \end{subfigure}
    \begin{subfigure}[t]{0.32\textwidth}
        \centering
        \includegraphics[width=\linewidth]{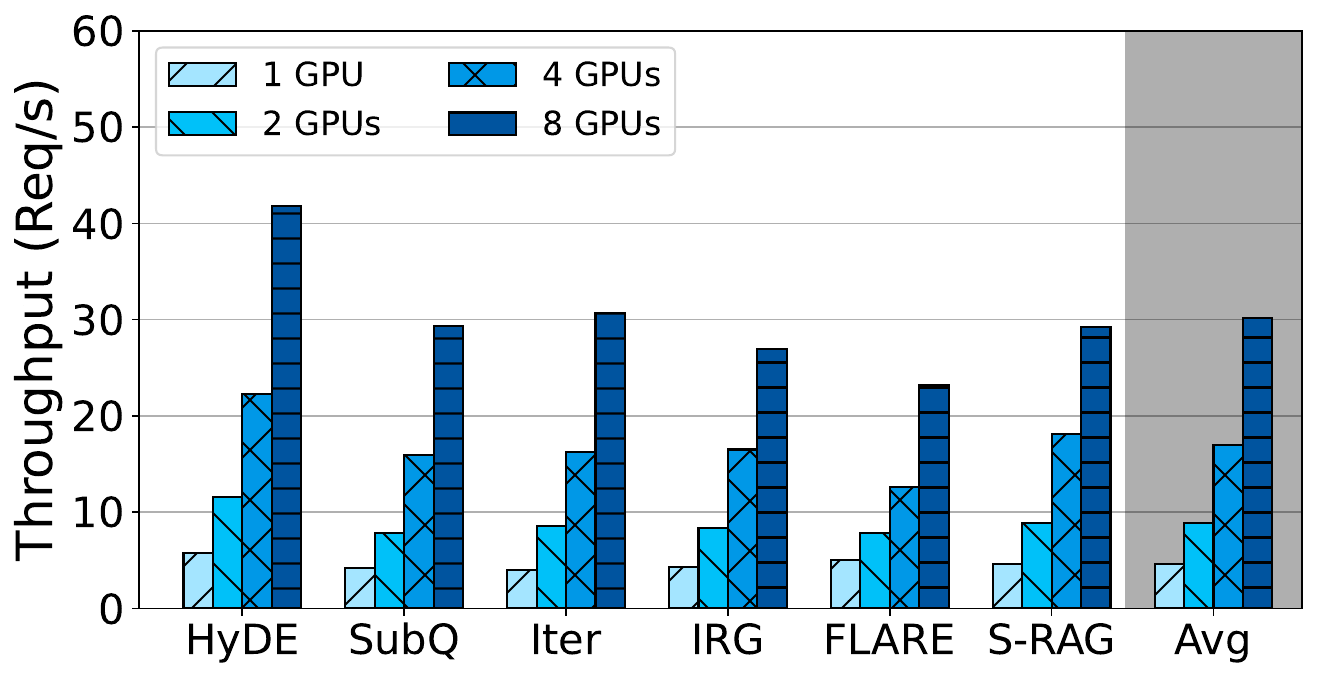}
        \vspace{-1.3em}
        \caption{End-to-end throughput on HotpotQA.}
    \end{subfigure}
    \begin{subfigure}[t]{0.32\textwidth}
        \centering
        \includegraphics[width=\linewidth]{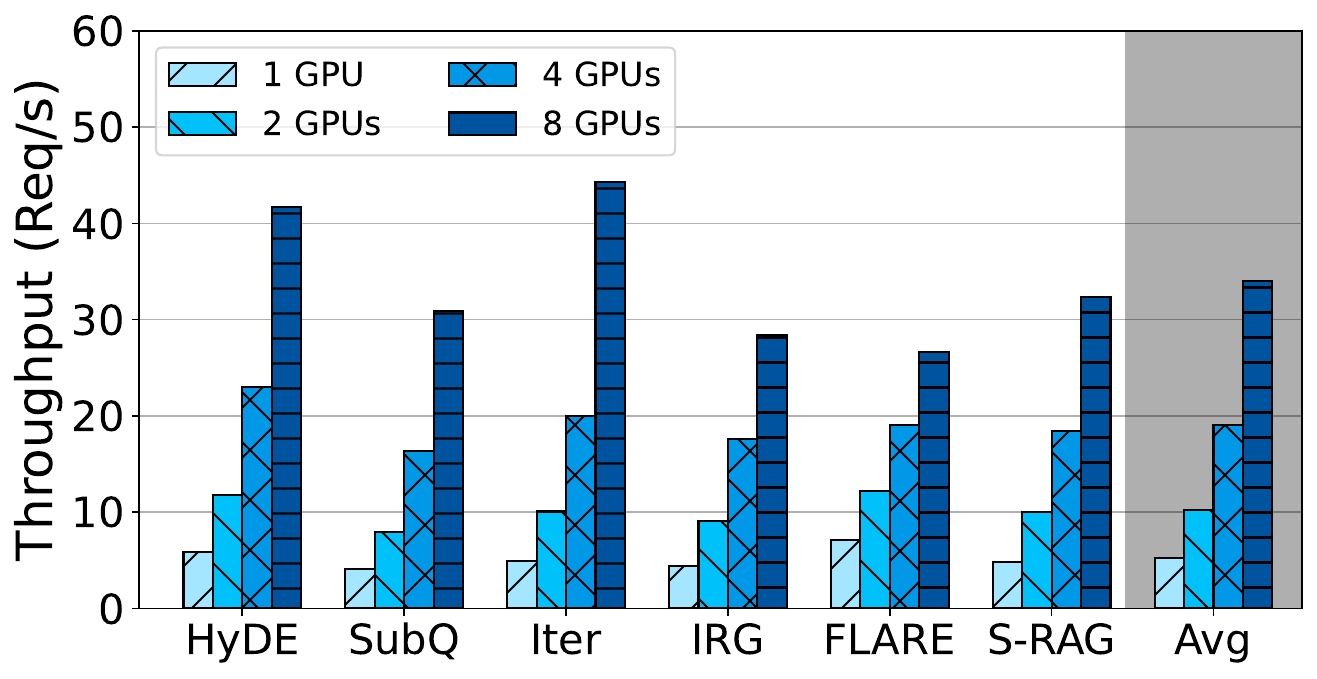}
        \vspace{-1.3em}
        \caption{End-to-end throughput on TriviaQA.}
    \end{subfigure}
    \vspace{-0.6em}
    \caption{Throughput scaling of \sys across multiple \gpuhuge GPUs. Global batch fixed at 128; micro-batch fixed at 4.
    }
    \label{fig:result-throughput-multi-gpu}
    \vspace{-1em}
\end{figure*}

\pgheading{Max GPU memory for retrieval}
We set a maximum GPU memory limit for prefetching in each configuration.
For \serverone{} and \servertwo{}, we allocated 12\,GB and 24\,GB, respectively.
For \desktop{}, we allocated up to 10\,GB and 3.75\,GB for the 3B and 8B models, respectively.

\pgheading{Cache Setup}
The memory allocated for retrieval is shared between prefetching and caching. In our evaluation, we set the cache proportion to 50\% (\ie the cache can use at most half of this memory). Since the benefit of caching is negligible on a single GPU (demonstrated in the ablation study), we enabled it only for the multi-GPU experiments on \servertwo.

\subsection{Evaluation Results}

\pgheading{Single-query latency on \gpusmall}
We evaluated the end-to-end RAG latency for a single query on \desktop with an \gpusmall GPU, representing the typical local usage. Figure~\ref{fig:result-speedup-rtx4090} shows the latency reduction of \sys across three datasets and two LLMs (\llamasmall and \llama).
As Figure \ref{fig:result-speedup-rtx4090} shows, \sys consistently outperforms the CPU-offload baseline across all evaluated configurations. 
With \llamasmall, \sys achieves average speedups of 1.55$\times$, 1.54$\times$, and 1.49$\times$ on NQ, HotpotQA, and TriviaQA, respectively. 

Among the test pipelines, the best speedup of 2.11$\times$ is achieved in the Iter-RetGen pipeline on HotpotQA.
It's because Iter-RetGen involves frequent retrieval operations and has generally short LLM outputs, which enhances the relative impact of retrieval acceleration.
Another notable improvement is in the SubQuestion pipeline, where \sys achieves approximately 1.85$\times$ speedup across all datasets. 
This pipeline uses LLM-generated sub-questions and performs batched retrievals of 3 to 4 queries in its retrieval stage. 
CPU-based retrieval suffers from limited parallelism in such scenarios, but \sys efficiently utilizes GPU parallelism, significantly enhancing performance.

When deploying \llama, the speedups from \sys{} are slightly smaller than with \llamasmall{}, mainly because the larger LLM incurs higher inference latency and leaves less GPU memory available for prefetching. Nevertheless, \sys{} still delivers an average speedup of about 1.3$\times$ across datasets, with a peak improvement of 1.82$\times$ for Iter-RetGen on HotpotQA.
Notably, these gains are achieved with only 3.75\,GB of remaining GPU memory after accounting for \llama (16\,GB), the embedding model (1\,GB), the KV cache, and other miscellaneous tensors. This result highlights \sys{}'s strong ability to accelerate RAG inference even under tight GPU memory constraints.

\begin{figure}[tb]
\centering
    \includegraphics[width=.9\columnwidth]{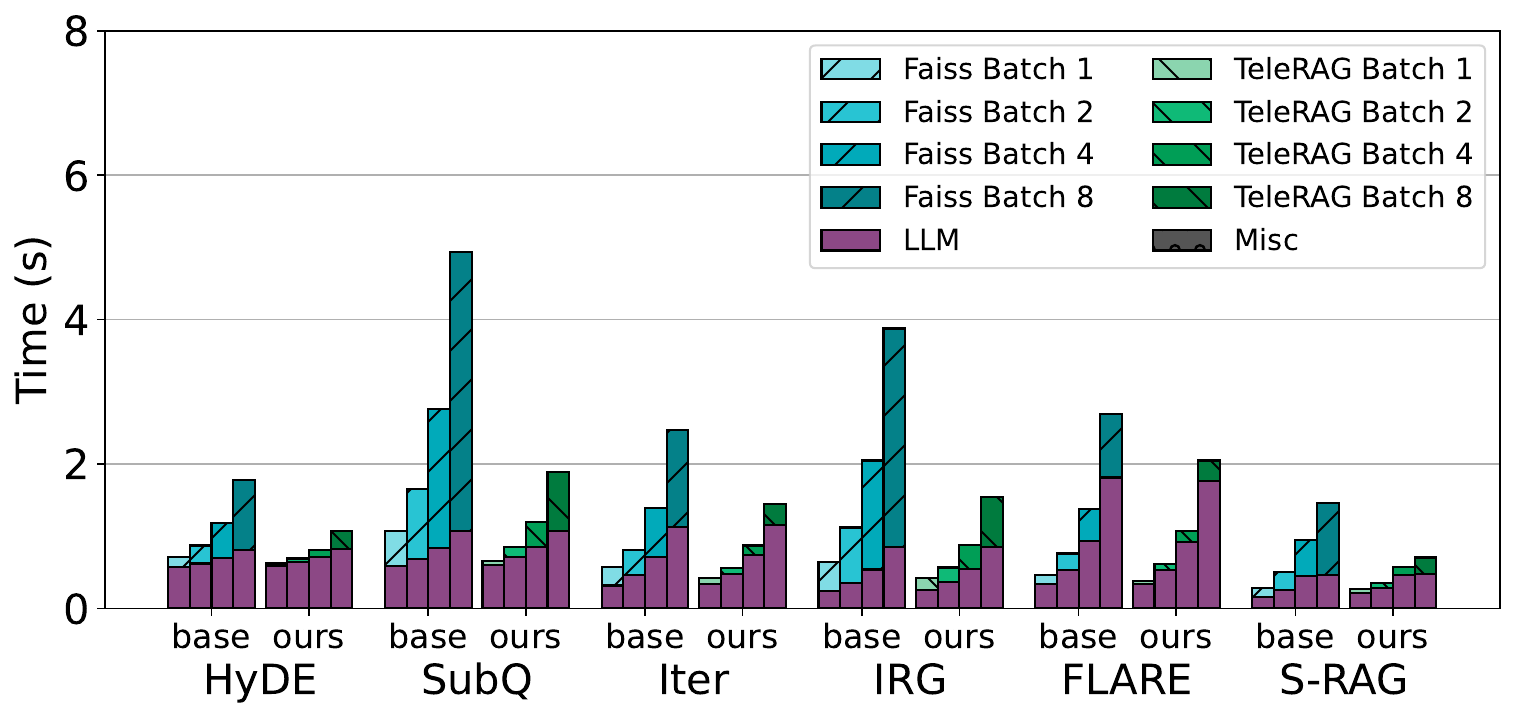}
    \vspace{-1em}
    \caption{Latency breakdown for \llama on NQ with an \gpularge GPU at different batch sizes. \nprobe is 256.}
    \label{fig:result-latency-breakdown-llama3-8b}
    \vspace{-0.8em}
\end{figure}

\begin{figure*}[tb]
    \centering
    \includegraphics[width=0.98\textwidth]{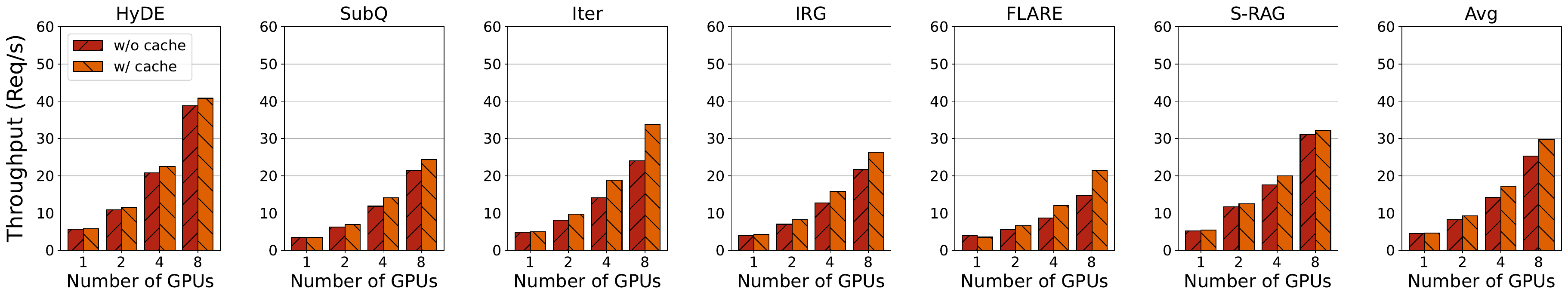}
    \vspace{-0.8em}
    \caption{Throughput of \sys on the NQ dataset with different numbers of \gpuhuge GPUs (with and without cache).}
    \label{fig:result-multi-gpu-cache-comparison}
    \vspace{-0.8em}
\end{figure*}

\begin{table}[tb]
    \centering 
    \resizebox{\columnwidth}{!}{
    \begin{tabular}{c|cc|cc|cc}
        \toprule
        \multirow{2}*{Pipeline}	& \multicolumn{2}{c|}{H100 (Mst-22B)} & \multicolumn{2}{c|}{H100 (Llm3-8B)} & \multicolumn{2}{c}{RTX4090 (Llm3-3B)}	\\
        & Budget & Hit Rate & Budget & Hit Rate & Budget & Hit Rate\\
        \midrule
        HyDE & 18 GB & \cellcolor{green9} $95.1\%$ & 10 GB & \cellcolor{green9} $93.2\%$ & 7 GB & \cellcolor{green9} $87.3\%$ \\
        SubQ & 18 GB & \cellcolor{green8} $85.2\%$ & 8 GB & \cellcolor{green8} $79.1\%$ & 7 GB & \cellcolor{green7} $76.4\%$ \\
        Iter & 7 GB & \cellcolor{green9} $97.4\%$ & 5 GB & \cellcolor{green9} $93.7\%$ & 3 GB & \cellcolor{green8} $72.6\%$ \\
        IRG & 12 GB & \cellcolor{green6} $65.4\%$ & 4 GB & \cellcolor{green6} $59.1\%$ & 2.5 GB & \cellcolor{green5} $45.1\%$ \\
        FLARE & 12 GB & \cellcolor{green9} $94.9\%$ & 6 GB & \cellcolor{green8} $87.8\%$ & 3 GB & \cellcolor{green6} $62.9\%$ \\
        S-RAG & 4.5 GB & \cellcolor{green9} $95.0\%$ & 3 GB & \cellcolor{green7} $72.6\%$ & 1.25 GB & \cellcolor{green3} $31.0\%$ \\
        \bottomrule
    \end{tabular}
    }
    \includegraphics[width=0.8\columnwidth]{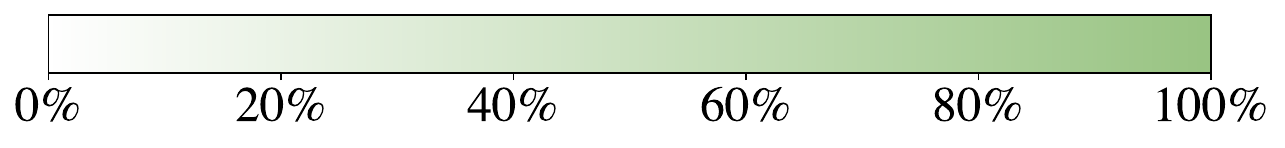}
    \vspace{-1em}
    \caption{The prefetch budget and corresponding averaged cluster hit rate for each pipeline and hardware setup on NQ dataset. The target retrieval \nprobe is 256.}
    \label{tab:result-hitrate}
    \vspace{-1em}
\end{table}

\pgheading{Multi-query throughput on \gpularge} 
To examine \sys{}'s performance on batched inference,
we evaluated the end-to-end throughput on \serverone (\gpularge) using batch sizes 1, 2, 4, and 8.
The results of six RAG pipelines with \llama \cameraready{and \mistral{}} are presented in Figure~\ref{fig:result-throughput-h100}.

As shown in Figure~\ref{fig:result-throughput-h100}, \sys consistently outperforms the Faiss baseline across all pipelines and batch sizes.
At batch size 1 (equivalent to the single-query setting), \sys{} delivers an average throughput increase of 1.32$\times$ and 1.15$\times$ for \llama and \mistral{} over the Faiss baseline.
As the batch size increases, \sys{}'s performance gains continue to grow.
For batch sizes 2, 4, and 8, \sys delivers average throughput increases of 1.55$\times$, 1.78$\times$ and 1.98$\times$ for \llama{}, and 1.26$\times$, 1.38$\times$ and 1.49$\times$ for \mistral{} over Faiss.
These results show that \sys effectively utilizes the parallel compute of the GPU without overwhelming its memory, while the CPU alternative fails to scale with a larger batch size.

\pgheading{Multi-GPU throughput} 
We evaluate \sys{}'s scalability on \servertwo{} (\gpuhuge), a multi-GPU system.
Figure~\ref{fig:result-throughput-multi-gpu} reports throughput for a global batch size of 128 with a micro-batch size of 4 across 1--8 GPUs, with both the prefetching and cache schedulers enabled. 
\sys{} scales well with the number of GPUs: on NQ, compared to the single-GPU case, the average speedups are 2.0$\times$, 3.8$\times$, and 6.5$\times$ on 2, 4, and 8 GPUs, respectively.
The mild sub-linear scaling at higher GPU counts is attributed to execution time variance across micro-batches---specifically, a long tail of higher-latency batches that create load imbalance.
These results still demonstrate that the \lookah{} technique in \sys{} can scale effectively.

\subsection{Analysis and Sensitivity Study}
\label{sec:evaluation-analysis}

\pgheading{Latency breakdown}
We further show the latency breakdown of running RAG pipelines with \llama on a single \gpularge GPU at different batch sizes in Figure~\ref{fig:result-latency-breakdown-llama3-8b}.
From Figure~\ref{fig:result-latency-breakdown-llama3-8b}, we can observe that LLM latency grows sub-linearly with larger batch sizes. However, the latency for Faiss retrieval on CPU grows linearly with the batch size, dominating the overall latency when the batch size is large.
These results echo our findings in Figure~\ref{fig:result-throughput-h100}, and show the limited scalability of CPU retrieval in serving scenarios.
In contrast, \sys significantly accelerates across all batch sizes and achieves a higher speedup from 1.3$\times$ to 2.0$\times$ when the batch size increases from 1 to 8.
\cameraready{The slight difference in the LLM latency between Faiss and \sys{} is because the PyTorch copy operation we use for prefetching uses some of the GPU's stream multiprocessors.}

\pgheading{Prefetch budgets, cluster hit rates \cameraready{and failure cases}}
Table~\ref{tab:result-hitrate} shows the prefetch budgets we set with the profile-guided approach on \gpusmall and \gpularge for NQ.
It also presents the average cluster hit rate achieved with this prefetch budget.
From the table, we can see that \sys generally achieves a high cluster hit rate (${>}{50\%}$) when it has a large prefetching budget.
For cases where the budget is less than 2~GB, we observe a relatively low hit rate (${<}{50\%}$), limiting the benefits of reducing the CPU's search workloads.
However, as observed from Figure~\ref{fig:result-speedup-rtx4090}, \sys achieves from 1.2$\times$ to 1.6$\times$ end-to-end speedups for these pipelines, thanks to the combined benefit of reducing CPU workload and utilizing the GPU to perform sorting on similarity distances.

\cameraready{We also conducted analysis on failure cases. Although failure cases can occur (\eg when the rewrite substantially shifts embedding vectors), it is relatively rare. Even with the lowest prefetch budget (3.75\,GB for \gpusmall with \llama) in our evaluation, the prefetching failures ($<$5\% hit rate) occurred only on the FLARE pipeline with a small number of instances (3, 14, and 9 out of 1024 samples for NQ, HotpotQA, and TriviaQA, respectively).}

\pgheading{Ablation on cache}
Figure~\ref{fig:result-multi-gpu-cache-comparison} illustrates the throughput improvements enabled by caching on the NQ dataset.
On average, caching boosts throughput by 2\%, 12\%, 21\%, and 18\% with 1, 2, 4, and 8 GPUs, respectively.
The benefit on a single GPU is marginal because only a small cache space is used, and the limited cache capacity struggles to accommodate diverse requests.
However, as the number of GPUs and overall cache space increases, the cache-aware scheduler effectively assigns micro-batches to the appropriate GPUs to maximize cache overlap, thereby improving prefetching hit rates and yielding significant throughput gains.

\begin{figure}[tb]
\centering
    \includegraphics[width=.9\columnwidth]{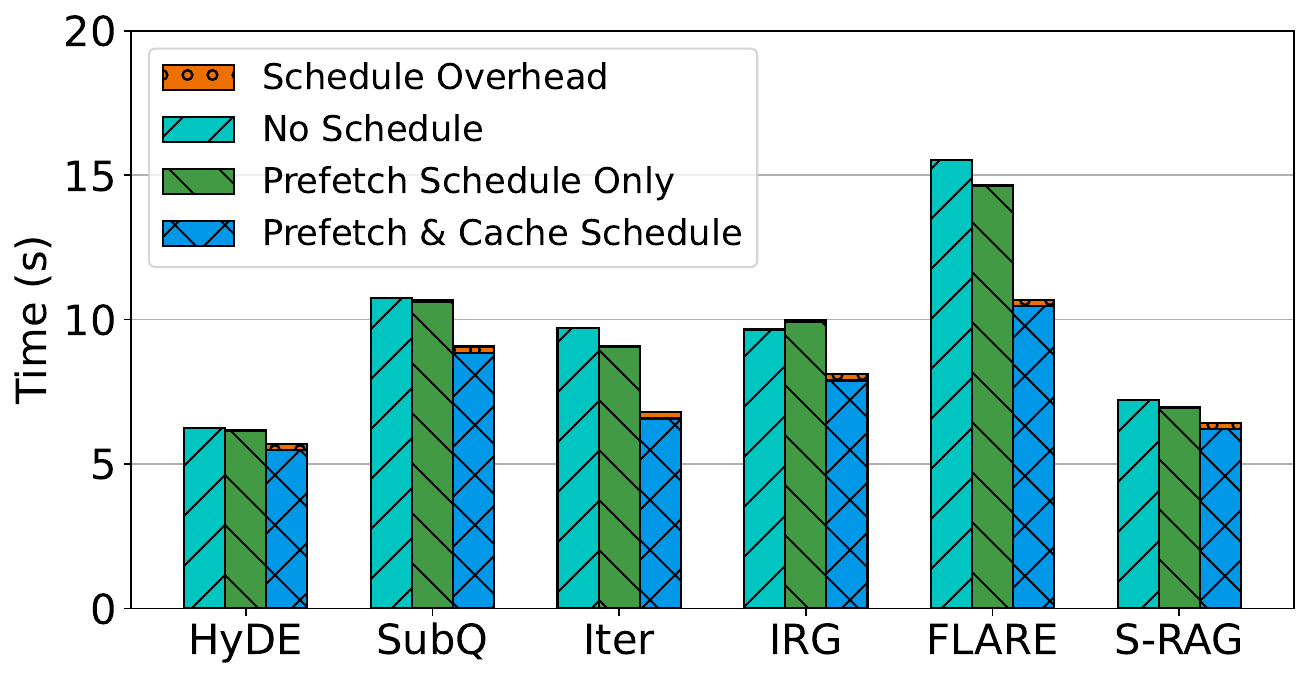}
    \vspace{-0.8em}
    \caption{Comparison of end-to-end latency for prefetching and cache-aware schedulers on 4 \gpuhuge GPUs.
    The overhead bars of prefetch schedule only are too short to be visible.
    }
    \label{fig:result-latency-breakdown-schedulers}
    \vspace{-1.5em}
\end{figure}

\pgheading{Analysis on scheduling overhead}
Figure~\ref{fig:result-latency-breakdown-schedulers} analyzes the benefits and overheads of the prefetching and cache-aware schedulers.
It reports the end-to-end latency for a global batch of 128 queries with a micro-batch size of 4 on 4 \gpuhuge{} GPUs, comparing configurations that enable both schedulers, only the prefetching scheduler, or neither.
As shown in Figure~\ref{fig:result-latency-breakdown-schedulers}, the prefetching scheduler incurs only a minimal overhead of approximately 37\,ms (thus almost invisible in the figure), while the cache-aware scheduler adds a modest overhead of about \cameraready{180\,ms}.
Overall, both schedulers effectively reduce end-to-end latency in most cases, with minimal additional cost.

\section{Related Work} \label{sec:related}

\pgheading{Efficient RAG methods}
Prior works have accelerated RAG by caching the LLM's KV cache from retrieved documents~\cite{jin2024ragcache, lu2024turborag, yao2024cacheblend}.
However, these approaches primarily reduce prefill latency, failing to address the retrieval and decode latencies which often dominate.
Other techniques, like speculative retrieval~\cite{zhang2024accelerating} or fine-grained pipelining~\cite{jiang2024piperag}, target per-token retrieval,
which is a different RAG paradigm and does not apply to the modular pipelines discussed in \S\ref{sec:background-rag}.
Unlike these approaches, \sys tackles the fundamental system challenges of long retrieval latency and large memory requirements for modular RAG.

\cameraready{
\pgheading{System optimizations for modular RAG}
There are several concurrent works to \sys that optimize system efficiency for modular RAG.
For increasing throughput, HedraRAG~\cite{hu2025hedrarag} leverages intra-request similarity and inter-query skew to accelerate retrieval with compute graph transformation, query reordering and dynamic placement between CPU and GPU;
RAGO~\cite{jiang2025rago} designs a schema for modular RAG and enables automatic performance analysis and optimizations on top of the schema;
Hermes~\cite{shen2025hermes} distributes the datastore across multiple CPU nodes and proposes efficient search algorithms to accelerate retrieval across nodes.
}

\cameraready{
For reducing memory, EdgeRAG~\cite{seemakhupt2024_edgerag} proposes to generate embeddings on-the-fly and offload embeddings to disk for the IVF index;
LEANN~\cite{wang2025leann} designs efficient embedding recomputation for the HNSW~\cite{malkov_hnsw} index.
In contrast to these works, \sys achieves both acceleration and GPU memory saving for retrieval with the IVF index.
}

\pgheading{Systems for compound LLM applications}
Apart from RAG, there is a growing interest in compound or agentic LLM applications, where multiple LLM calls and other applications are combined to serve complex functionalities~\cite{compound-ai-blog,sigarch,wang2024survey}. 
AI Metropolis~\cite{xie2024ai} accelerates LLM-based multi-agent simulations with out-of-order execution.
RAG is a specific type of application in this broader direction, and we propose systems techniques to optimize its execution latency, focusing on the characteristics of retrieval workload.

\pgheading{Vector index}
A separate line of work focuses on optimizing the vector index itself.
This includes hardware-specific acceleration (GPU~\cite{johnson2019billion}, FPGA~\cite{jiang2023co})
and hybrid memory-disk systems to scale beyond RAM (DiskANN~\cite{jayaram2019diskann}, SPANN~\cite{chen2021spann}).
More recent systems scale graph-based indexes beyond GPU memory~\cite{karthik2025bang, zhang2024fast}.
These methods often require significant algorithm modifications or remain bottlenecked by CPU--GPU bandwidth.
\sys is complementary, optimizing at the system-level by hiding transfer latency within the RAG application's context,
without altering the underlying IVF algorithm.

\section{Discussion}

\camerareadytwo{
\pgheading{Applicability beyond IVF indices}
\sys{} is designed for IVF-style retrieval indices, which are widely used in production vector search systems~\cite{douze2024faiss, milvus, pgvector}. Here, we briefly discuss how its design extends to two other popular index families: Locality-Sensitive Hashing (LSH)~\cite{indyk1998approximate,gionis1999similarity,datar2004locality} and Hierarchical Navigable Small World (HNSW)~\cite{malkov_hnsw}.
}

\camerareadytwo{
LSH partitions vectors into hash buckets and probes a small number of buckets at query time. Since this access pattern resembles IVF list probing, \sys{} can be extended naturally by treating hash buckets as the prefetch unit.
}

\camerareadytwo{
HNSW, by contrast, relies on graph traversal and does not naturally expose explicit prefetchable units, making it less naturally compatible with \sys{}. Nevertheless, the core ideas of \sys{} still apply to hybrid index structures such as IVF-HNSW in FAISS~\cite{nonexhaustive_faiss}. In these indices, HNSW is used for coarse-grained routing to identify relevant inverted lists. Because semantically similar queries often traverse similar graph paths or reach the same centroids, \sys{} can use the HNSW traversal of a pre-retrieval query to anticipate and prefetch the corresponding inverted lists before the retrieval query is issued. This preserves the efficiency of graph-based routing while retaining the prefetchability of the inverted-list structure.
}


\cameraready{
\pgheading{Applicability of emerging hardware}
\sys{} targets the traditional GPU computing platform, where CPU and GPU have separate memory space and are connected through PCIe.
On newer high-bandwidth systems such as NVIDIA Grace~\cite{nvidiagrace}, our design can leverage the increased bandwidth to prefetch more clusters, thereby improving hit rates.
For emerging unified memory platforms such as Apple M-series or NVIDIA Grace-Hopper, \sys is less useful as we can do GPU-only retrieval on those.
However, these systems currently still suffer from limited memory capacity (\eg Apple M-series) or are not as available as the traditional commodity DDR + discrete HBM servers.
}

\cameraready{
\pgheading{Memory allocation tradeoff}
\sys{} is designed so that retrieval does not compete with the LLM for KV-cache capacity. We reserve GPU memory for the LLM first and use only the remaining memory for retrieval data. Accordingly, we enforce a hard cap on retrieval memory, as described in \S\ref{sec:exp_setups}.
}

\cameraready{
In our evaluated workloads, this policy leaves sufficient KV-cache headroom and does not reduce LLM batch size or degrade LLM performance.
In practice, the retrieval memory cap should be chosen only after reserving enough memory for the target model's KV cache at the desired batch size and sequence length; retrieval should use only the remaining GPU memory.
}

\cameraready{
\pgheading{Multi-GPU and multi-node parallelism}
In \sys{}, data parallelism is the default strategy to extend to multi-GPU and multi-node deployments, as it preserves the overlap between LLM generation and retrieval as in the single-GPU design.
Here, we discuss other potential parallelism patterns.
}

\cameraready{
Another option to parallelize retrieval is to shard the datastore across GPUs for distributed GPU-resident retrieval.
This is most effective when the vector index fits in aggregate GPU memory without materially reducing the memory available to the LLM, especially the KV cache.
When this condition does not hold, \sys{} is useful to reduce the retrieval memory usage to a smaller predicted set. 
}

\cameraready{
Furthermore, \sys{}'s retrieval and generation can be pipelined across nodes. A retrieval tier can start prefetching as soon as it receives the pre-retrieval query or predicted clusters, overlapping transfer with the pre-retrieval generation on another tier. 
While pipeline bubbles may arise if the stages are imbalanced, \lookah still improves efficiency by shortening the retrieval critical path and reducing GPU memory requirements for retrieval.
We leave the exploration of different parallelisms for future work.
}

%
%

\section{Conclusion}
\label{sec:conclusion}
In this paper, we introduced \sys, an inference system that improves RAG pipeline latency and throughput while imposing minimal GPU memory requirements.
\sys achieves significant acceleration by employing \textit{lookahead retrieval}, which hides CPU--GPU data transfer latency by overlapping it with pre-retrieval generation, and by supporting efficient multi-GPU inference through specialized scheduling.
Our evaluation shows that \sys significantly improves performance compared to existing state-of-the-art solutions and scales effectively with increased hardware resources.

\section*{Acknowledgments}

We thank the reviewers and our shepherd for their helpful comments and suggestions.
This work is supported in part by PRISM, one of the seven centers in JUMP~2.0, a Semiconductor Research Corporation (SRC) program sponsored by DARPA, as well as generous donations from NVIDIA, AMD, Intel, and Arm.


\bibliography{reference}

@article{gao2023retrieval_rag_survey,
  title={Retrieval-augmented generation for large language models: A survey},
  author={Gao, Yunfan and Xiong, Yun and Gao, Xinyu and Jia, Kangxiang and Pan, Jinliu and Bi, Yuxi and Dai, Yi and Sun, Jiawei and Wang, Haofen},
  journal={arXiv preprint arXiv:2312.10997},
  year={2023}
}

@inproceedings{fan2024_rag_survey,
  title={A Survey on RAG Meeting LLMs: Towards Retrieval-Augmented Large Language Models},
  author={Wenqi Fan and Yujuan Ding and Liang-bo Ning and Shijie Wang and Hengyun Li and Dawei Yin and Tat-Seng Chua and Qing Li},
  booktitle={Knowledge Discovery and Data Mining},
  year={2024},
  url={https://api.semanticscholar.org/CorpusID:269740933}
}

@article{asai2024reliable,
  title={Reliable, adaptable, and attributable language models with retrieval},
  author={Asai, Akari and Zhong, Zexuan and Chen, Danqi and Koh, Pang Wei and Zettlemoyer, Luke and Hajishirzi, Hannaneh and Yih, Wen-tau},
  journal={arXiv preprint arXiv:2403.03187},
  year={2024}
}

@software{llamaindex,
    author = {Liu, Jerry},
    doi = {10.5281/zenodo.1234},
    month = {11},
    title = {{LlamaIndex}},
    howpublished = {\url{https://github.com/jerryjliu/llama_index}},
    year = {2022}
}

@software{sub_question_llamaindex,
    author = {Liu, Jerry},
    month = {11},
    title = {{Sub Question Query Engine LlamaIndex}},
    howpublished = {\url{https://developers.llamaindex.ai/python/examples/query_engine/sub_question_query_engine/}},
    year = {2022}
}

@software{multistep_llamaindex,
    author = {Liu, Jerry},
    month = {11},
    title = {{MultiStep Query Engine LlamaIndex}},
    howpublished = {\url{https://developers.llamaindex.ai/python/examples/workflow/multi_step_query_engine/}},
    year = {2022}
}

@article{openscholar,
  title={OpenScholar: Synthesizing Scientific Literature with Retrieval-augmented LMs},
  author={Asai, Akari and He, Jacqueline and Shao, Rulin and Shi, Weijia and Singh, Amanpreet and Chang, Joseph Chee and Lo, Kyle and Soldaini, Luca and Feldman, Sergey and D'arcy, Mike and others},
  journal={arXiv preprint arXiv:2411.14199},
  year={2024}
}

@misc{llama3modelcard,
    title={Llama 3 Model Card},
    author={AI@Meta},
    year={2024},
    howpublished = {\url{https://github.com/meta-llama/llama3/blob/main/MODEL_CARD.md}},
    note         = {Accessed: 2026-04-19}
}

@inproceedings{karpukhin2020dense,
    title = "Dense Passage Retrieval for Open-Domain Question Answering",
    author = "Karpukhin, Vladimir and Oguz, Barlas and Min, Sewon and Lewis, Patrick and Wu, Ledell and Edunov, Sergey and Chen, Danqi and Yih, Wen-tau",
    booktitle = "Proceedings of the 2020 Conference on Empirical Methods in Natural Language Processing (EMNLP)",
    year = "2020",
    howpublished = {\url{https://www.aclweb.org/anthology/2020.emnlp-main.550}},
}

@article{johnson2019billion,
  title={Billion-scale similarity search with {GPUs}},
  author={Johnson, Jeff and Douze, Matthijs and J{\'e}gou, Herv{\'e}},
  journal={IEEE Transactions on Big Data},
  volume={7},
  number={3},
  pages={535--547},
  year={2019},
  publisher={IEEE}
}

@inproceedings{ma2023query,
  title={Query Rewriting in Retrieval-Augmented Large Language Models},
  author={Ma, Xinbei and Gong, Yeyun and He, Pengcheng and Duan, Nan and others},
  booktitle={The 2023 Conference on Empirical Methods in Natural Language Processing},
  year={2023}
}

@misc{vonage_reducing_rag_latency,
  title        = {Reducing RAG Pipeline Latency for Real-Time Voice Conversations},
  author       = {Binoy Chemmagate},
  howpublished = {\url{https://developer.vonage.com/en/blog/reducing-rag-pipeline-latency-for-real-time-voice-conversations}},
  year         = {2024},
  note         = {Accessed: 2026-04-19}
}

@article{akkiraju2024facts,
  title={FACTS About Building Retrieval Augmented Generation-based Chatbots},
  author={Akkiraju, Rama and Xu, Anbang and Bora, Deepak and Yu, Tan and An, Lu and Seth, Vishal and Shukla, Aaditya and Gundecha, Pritam and Mehta, Hridhay and Jha, Ashwin and others},
  journal={arXiv preprint arXiv:2407.07858},
  year={2024}
}

@misc{ontinue_latency_ion_iq_chatbot,
  title={Enhancing User Experience by Overcoming Latency in the ION IQ Chatbot},
    author={Kuan Tung},
  howpublished = {\url{https://www.ontinue.com/resource/enhancing-user-experience-by-overcoming-latency-in-the-ion-iq-chatbot/}},
  year         = {2024},
  note         = {Accessed: 2026-04-19}
}

@misc{harchworks_rag_financial,
  title={RAG in Financial Services: Use-Cases, Impact, \& Solutions},
    author={Melissa Malec},
  howpublished = {\url{https://hatchworks.com/blog/gen-ai/rag-for-financial-services/}},
  year         = {2024},
  note         = {Accessed: 2026-04-19}
}

@misc{myscale_rag_trading,
  title={4 Key Benefits of RAG Algorithmic Trading in Financial Markets},
    author={MyScale},
  howpublished = {\url{https://myscale.com/blog/benefits-rag-algorithmic-trading-financial-markets/}},
  year         = {2024},
  note         = {Accessed: 2026-04-19}
}

@misc{apollo_medlm_rag,
  title={How Apollo 24|7 leverages MedLM with RAG to revolutionize healthcare},
    author={Abdussamad GM and Gopala Dhar},
  howpublished = {\url{https://cloud.google.com/blog/products/ai-machine-learning/how-apollo-247-leverages-medlm-with-rag-to-revolutionize-healthcare}},
  year         = {2024},
  note         = {Accessed: 2026-04-19}
}

@article{klang2024assessing,
  title={Assessing Retrieval-Augmented Large Language Model Performance in Emergency Department ICD-10-CM Coding Compared to Human Coders},
  author={Klang, Eyal and Tessler, Idit and Apakama, Donald U and Abbott, Ethan and Glicksberg, Benjamin S and Arnold, Monique and Moses, Akini and Sakhuja, Ankit and Soroush, Ali and Charney, Alexander W and others},
  journal={medRxiv},
  year={2024}
}

@misc{cohere-rerank,
      title={Say Hello to Precision: How Rerankers and Embeddings Boost Search}, 
      author={David Stewart and Jamie Linsdell},
      howpublished={\url{https://cohere.com/blog/say-hello-to-precision-how-rerankers-and-embeddings-boost-search}},
      year={2024},
      note={Accessed: 2026-04-19}
}

@misc{advanced-rag-overview,
      title={Advanced RAG Techniques: an Illustrated Overview}, 
      author={Ivan Ilin},
      howpublished={\url{https://pub.towardsai.net/advanced-rag-techniques-an-illustrated-overview-04d193d8fec6}},
      year={2023},
      note={Accessed: 2026-04-19}
}

@misc{databricks-blog,
      title={RAG (Retrieval Augmented Generation) on Databricks}, 
      author={Databricks},
      howpublished={\url{https://docs.databricks.com/en/generative-ai/retrieval-augmented-generation.html}},
      year={2024},
      note={Accessed: 2026-04-19}
}

@misc{azure-blog,
      title={Advanced RAG with Azure AI Search and LlamaIndex}, 
      author={Khye Wei},
      howpublished={\url{https://techcommunity.microsoft.com/t5/ai-azure-ai-services-blog/advanced-rag-with-azure-ai-search-and-llamaindex/ba-p/4115007}},
      year={2024},
      note={Accessed: 2026-04-19}
}

@misc{zilliz-blog,
      title={How to Select Index Parameters for IVF Index}, 
      author={Zilliz},
      howpublished={\url{https://zilliz.com/blog/select-index-parameters-ivf-index}},
      year={2020},
      note={Accessed: 2026-04-19}
}

@misc{modular-rag-survey,
      title={Modular RAG and RAG Flow: Part II}, 
      author={Yunfan Gao},
      howpublished={\url{https://medium.com/@yufan1602/modular-rag-and-rag-flow-part-ii-77b62bf8a5d3}},
      year={2024},
      note={Accessed: 2026-04-19}
}

@article{zheng2023take,
  title={Take a step back: Evoking reasoning via abstraction in large language models},
  author={Zheng, Huaixiu Steven and Mishra, Swaroop and Chen, Xinyun and Cheng, Heng-Tze and Chi, Ed H and Le, Quoc V and Zhou, Denny},
  journal={arXiv preprint arXiv:2310.06117},
  year={2023}
}

@article{jagerman2023query,
  title={Query expansion by prompting large language models},
  author={Jagerman, Rolf and Zhuang, Honglei and Qin, Zhen and Wang, Xuanhui and Bendersky, Michael},
  journal={arXiv preprint arXiv:2305.03653},
  year={2023}
}

@inproceedings{gao2022precise_hyde,
  title={Precise Zero-Shot Dense Retrieval without Relevance Labels},
  author={Gao, Luyu and Ma, Xueguang and Lin, Jimmy and Callan, Jamie},
  booktitle={Proceedings of the 61st Annual Meeting of the Association for Computational Linguistics (Volume 1: Long Papers)},
  pages={1762--1777},
  year={2023}
}

@article{jiang2023longllmlingua,
  title={Longllmlingua: Accelerating and enhancing llms in long context scenarios via prompt compression},
  author={Jiang, Huiqiang and Wu, Qianhui and Luo, Xufang and Li, Dongsheng and Lin, Chin-Yew and Yang, Yuqing and Qiu, Lili},
  journal={arXiv preprint arXiv:2310.06839},
  year={2023}
}

@inproceedings{asai2023self,
  title={Self-RAG: Learning to Retrieve, Generate, and Critique through Self-Reflection},
  author={Asai, Akari and Wu, Zeqiu and Wang, Yizhong and Sil, Avirup and Hajishirzi, Hannaneh},
  booktitle={The Twelfth International Conference on Learning Representations},
  year={2023}
}

@misc{selfrag_code,
      title={Original implementation of SELF-RAG: Learning to Retrieve, Generate and Critique through self-reflection}, 
      author={Asai, Akari and Wu, Zeqiu and Wang, Yizhong and Sil, Avirup and Hajishirzi, Hannaneh},
      howpublished={\url{https://github.com/AkariAsai/self-rag}},
      year={2024}
}

@article{jiang2023flare,
      title={Active Retrieval Augmented Generation}, 
      author={Zhengbao Jiang and Frank F. Xu and Luyu Gao and Zhiqing Sun and Qian Liu and Jane Dwivedi-Yu and Yiming Yang and Jamie Callan and Graham Neubig},
      year={2023},
      eprint={2305.06983},
      archivePrefix={arXiv},
      primaryClass={cs.CL}
}

@inproceedings{ye2023enhancing,
  title={Enhancing Conversational Search: Large Language Model-Aided Informative Query Rewriting},
  author={Ye, Fanghua and Fang, Meng and Li, Shenghui and Yilmaz, Emine},
  booktitle={The 2023 Conference on Empirical Methods in Natural Language Processing},
  year={2023}
}

@inproceedings{zhuang2023open,
  title={Open-source Large Language Models are Strong Zero-shot Query Likelihood Models for Document Ranking},
  author={Zhuang, Shengyao and Liu, Bing and Koopman, Bevan and Zuccon, Guido},
  booktitle={Findings of the Association for Computational Linguistics: EMNLP 2023},
  pages={8807--8817},
  year={2023}
}

@inproceedings{zhou2022least,
  title={Least-to-Most Prompting Enables Complex Reasoning in Large Language Models},
  author={Zhou, Denny and Sch{\"a}rli, Nathanael and Hou, Le and Wei, Jason and Scales, Nathan and Wang, Xuezhi and Schuurmans, Dale and Cui, Claire and Bousquet, Olivier and Le, Quoc V and others},
  booktitle={The Eleventh International Conference on Learning Representations},
  year={2022}
}

@article{jin2024flashrag,
  title={FlashRAG: A Modular Toolkit for Efficient Retrieval-Augmented Generation Research},
  author={Jin, Jiajie and Zhu, Yutao and Yang, Xinyu and Zhang, Chenghao and Dou, Zhicheng},
  journal={arXiv preprint arXiv:2405.13576},
  year={2024}
}

@inproceedings{kim2023sure,
  title={SuRe: Improving Open-domain Question Answering of LLMs via Summarized Retrieval},
  author={Kim, Jaehyung and Nam, Jaehyun and Mo, Sangwoo and Park, Jongjin and Lee, Sang-Woo and Seo, Minjoon and Ha, Jung-Woo and Shin, Jinwoo},
  booktitle={The Twelfth International Conference on Learning Representations},
  year={2023}
}

@inproceedings{press2022measuring,
  title={Measuring and Narrowing the Compositionality Gap in Language Models},
  author={Press, Ofir and Zhang, Muru and Min, Sewon and Schmidt, Ludwig and Smith, Noah A and Lewis, Mike},
  booktitle={Findings of the Association for Computational Linguistics: EMNLP 2023},
  pages={5687--5711},
  year={2023}
}

@inproceedings{borgeaud2022improving,
  title={Improving language models by retrieving from trillions of tokens},
  author={Borgeaud, Sebastian and Mensch, Arthur and Hoffmann, Jordan and Cai, Trevor and Rutherford, Eliza and Millican, Katie and Van Den Driessche, George Bm and Lespiau, Jean-Baptiste and Damoc, Bogdan and Clark, Aidan and others},
  booktitle={International conference on machine learning},
  pages={2206--2240},
  year={2022},
  organization={PMLR}
}

@inproceedings{min2023silo,
  title={SILO Language Models: Isolating Legal Risk In a Nonparametric Datastore},
  author={Min, Sewon and Gururangan, Suchin and Wallace, Eric and Shi, Weijia and Hajishirzi, Hannaneh and Smith, Noah A and Zettlemoyer, Luke},
  booktitle={The Twelfth International Conference on Learning Representations},
  year={2023}
}

@inproceedings{mallen2022not,
  title={When Not to Trust Language Models: Investigating Effectiveness of Parametric and Non-Parametric Memories},
  author={Mallen, Alex Troy and Asai, Akari and Zhong, Victor and Das, Rajarshi and Khashabi, Daniel and Hajishirzi, Hannaneh},
  booktitle={The 61st Annual Meeting Of The Association For Computational Linguistics},
  year={2023}
}

@article{ram2023context,
  title={In-Context Retrieval-Augmented Language Models},
  author={Ram, Ori and Levine, Yoav and Dalmedigos, Itay and Muhlgay, Dor and Shashua, Amnon and Leyton-Brown, Kevin and Shoham, Yoav},
  journal={Transactions of the Association for Computational Linguistics},
  volume={11},
  pages={1316--1331},
  year={2023}
}

@inproceedings{khandelwal2019generalization,
  title={Generalization through Memorization: Nearest Neighbor Language Models},
  author={Khandelwal, Urvashi and Levy, Omer and Jurafsky, Dan and Zettlemoyer, Luke and Lewis, Mike},
  booktitle={International Conference on Learning Representations},
  year={2019}
}

@article{izacard2023atlas,
  title={Atlas: Few-shot learning with retrieval augmented language models},
  author={Izacard, Gautier and Lewis, Patrick and Lomeli, Maria and Hosseini, Lucas and Petroni, Fabio and Schick, Timo and Dwivedi-Yu, Jane and Joulin, Armand and Riedel, Sebastian and Grave, Edouard},
  journal={Journal of Machine Learning Research},
  volume={24},
  number={251},
  pages={1--43},
  year={2023}
}

@article{hardt2023test,
  title={Test-time training on nearest neighbors for large language models},
  author={Hardt, Moritz and Sun, Yu},
  journal={arXiv preprint arXiv:2305.18466},
  year={2023}
}

@inproceedings{zhang2024fast,
  title={Fast Vector Query Processing for Large Datasets Beyond {GPU} Memory with Reordered Pipelining},
  author={Zhang, Zili and Liu, Fangyue and Huang, Gang and Liu, Xuanzhe and Jin, Xin},
  booktitle={21st USENIX Symposium on Networked Systems Design and Implementation (NSDI 24)},
  pages={23--40},
  year={2024}
}

@misc{zheng2024_sglang,
      title={SGLang: Efficient Execution of Structured Language Model Programs}, 
      author={Lianmin Zheng and Liangsheng Yin and Zhiqiang Xie and Chuyue Sun and Jeff Huang and Cody Hao Yu and Shiyi Cao and Christos Kozyrakis and Ion Stoica and Joseph E. Gonzalez and Clark Barrett and Ying Sheng},
      year={2024},
      eprint={2312.07104},
      archivePrefix={arXiv},
      primaryClass={cs.AI},
      url={https://arxiv.org/abs/2312.07104}, 
}

@article{jin2024ragcache,
  title={RAGCache: Efficient Knowledge Caching for Retrieval-Augmented Generation},
  author={Jin, Chao and Zhang, Zili and Jiang, Xuanlin and Liu, Fangyue and Liu, Xin and Liu, Xuanzhe and Jin, Xin},
  journal={arXiv preprint arXiv:2404.12457},
  year={2024}
}

@misc{yao2024cacheblend,
      title={CacheBlend: Fast Large Language Model Serving for RAG with Cached Knowledge Fusion}, 
      author={Jiayi Yao and Hanchen Li and Yuhan Liu and Siddhant Ray and Yihua Cheng and Qizheng Zhang and Kuntai Du and Shan Lu and Junchen Jiang},
      year={2024},
      eprint={2405.16444},
      archivePrefix={arXiv},
      primaryClass={cs.LG},
      url={https://arxiv.org/abs/2405.16444}, 
}

@article{zhang2024accelerating,
  title={Accelerating retrieval-augmented language model serving with speculation},
  author={Zhang, Zhihao and Zhu, Alan and Yang, Lijie and Xu, Yihua and Li, Lanting and Phothilimthana, Phitchaya Mangpo and Jia, Zhihao},
  journal={arXiv preprint arXiv:2401.14021},
  year={2024}
}

@article{lewis2020retrieval,
  title={Retrieval-augmented generation for knowledge-intensive nlp tasks},
  author={Lewis, Patrick and Perez, Ethan and Piktus, Aleksandra and Petroni, Fabio and Karpukhin, Vladimir and Goyal, Naman and K{\"u}ttler, Heinrich and Lewis, Mike and Yih, Wen-tau and Rockt{\"a}schel, Tim and others},
  journal={Advances in Neural Information Processing Systems},
  volume={33},
  pages={9459--9474},
  year={2020}
}

@inproceedings{norouzi2013cartesian,
  title={Cartesian k-means},
  author={Norouzi, Mohammad and Fleet, David J},
  booktitle={Proceedings of the IEEE Conference on computer Vision and Pattern Recognition},
  pages={3017--3024},
  year={2013}
}

@article{jiang2024piperag,
  title={Piperag: Fast retrieval-augmented generation via algorithm-system co-design},
  author={Jiang, Wenqi and Zhang, Shuai and Han, Boran and Wang, Jie and Wang, Bernie and Kraska, Tim},
  journal={arXiv preprint arXiv:2403.05676},
  year={2024}
}

@inproceedings{quinn2025_iks,
author = {Quinn, Derrick and Nouri, Mohammad and Patel, Neel and Salihu, John and Salemi, Alireza and Lee, Sukhan and Zamani, Hamed and Alian, Mohammad},
title = {Accelerating Retrieval-Augmented Generation},
year = {2025},
isbn = {9798400706981},
publisher = {Association for Computing Machinery},
address = {New York, NY, USA},
url = {https://doi.org/10.1145/3669940.3707264},
doi = {10.1145/3669940.3707264},
booktitle = {Proceedings of the 30th ACM International Conference on Architectural Support for Programming Languages and Operating Systems, Volume 1},
pages = {15–32},
numpages = {18},
location = {Rotterdam, Netherlands},
series = {ASPLOS '25}
}

@misc{seemakhupt2024_edgerag,
      title={EdgeRAG: Online-Indexed RAG for Edge Devices}, 
      author={Korakit Seemakhupt and Sihang Liu and Samira Khan},
      year={2024},
      eprint={2412.21023},
      archivePrefix={arXiv},
      primaryClass={cs.LG},
      url={https://arxiv.org/abs/2412.21023}, 
}

@misc{wang2025leann,
      title={LEANN: A Low-Storage Vector Index},
      author={Yichuan Wang and Shu Liu and Zhifei Li and Yongji Wu and Ziming Mao and Yilong Zhao and Xiao Yan and Zhiying Xu and Yang Zhou and Ion Stoica and Sewon Min and Matei Zaharia and Joseph E. Gonzalez},
      year={2025},
      eprint={2506.08276},
      archivePrefix={arXiv},
      primaryClass={cs.DB},
      url={https://arxiv.org/abs/2506.08276},
}

@inproceedings{jiang2023co,
  title={Co-design hardware and algorithm for vector search},
  author={Jiang, Wenqi and Li, Shigang and Zhu, Yu and de Fine Licht, Johannes and He, Zhenhao and Shi, Runbin and Renggli, Cedric and Zhang, Shuai and Rekatsinas, Theodoros and Hoefler, Torsten and others},
  booktitle={Proceedings of the International Conference for High Performance Computing, Networking, Storage and Analysis},
  pages={1--15},
  year={2023}
}

@misc{rapidsai,
  title={Rapidsai/raft: RAFT contains fundamental widely-used algorithms and primitives for data science, Graph and machine learning.},
  howpublished={\url{https://github.com/rapidsai/raft}},
  journal={GitHub},
  publisher={Nvidia RAPIDS},
  author={Rapidsai},
  year={2022}
}

@inproceedings{yang2018hotpotqa,
  title={{HotpotQA}: A Dataset for Diverse, Explainable Multi-hop Question Answering},
  author={Yang, Zhilin and Qi, Peng and Zhang, Saizheng and Bengio, Yoshua and Cohen, William W. and Salakhutdinov, Ruslan and Manning, Christopher D.},
  booktitle={Conference on Empirical Methods in Natural Language Processing ({EMNLP})},
  year={2018}
}

@inproceedings{kwon2023efficient,
  title={Efficient Memory Management for Large Language Model Serving with PagedAttention},
  author={Woosuk Kwon and Zhuohan Li and Siyuan Zhuang and Ying Sheng and Lianmin Zheng and Cody Hao Yu and Joseph E. Gonzalez and Hao Zhang and Ion Stoica},
  booktitle={Proceedings of the ACM SIGOPS 29th Symposium on Operating Systems Principles},
  year={2023}
}

@inproceedings{shao2023enhancing,
  title={Enhancing Retrieval-Augmented Large Language Models with Iterative Retrieval-Generation Synergy},
  author={Shao, Zhihong and Gong, Yeyun and Shen, Yelong and Huang, Minlie and Duan, Nan and Chen, Weizhu},
  booktitle={Findings of the Association for Computational Linguistics: EMNLP 2023},
  pages={9248--9274},
  year={2023}
}

@inproceedings{peng2024large,
    author = {Peng, Wenjun and Li, Guiyang and Jiang, Yue and Wang, Zilong and Ou, Dan and Zeng, Xiaoyi and Xu, Derong and Xu, Tong and Chen, Enhong},
    title = {Large Language Model based Long-tail Query Rewriting in Taobao Search},
    year = {2024},
    booktitle = {Companion Proceedings of the ACM on Web Conference 2024},
}

@misc{sigarch,
    author={Berger, Emery and Zorn, Ben},
    title={AI Software Should be More Like Plain Old Software},
    howpublished={\url{https://www.sigarch.org/ai-software-should-be-more-like-plain-old-software/}},
    year={2024},
    note={Accessed: 2026-04-19}
}

@misc{compound-ai-blog,
  title={The Shift from Models to Compound AI Systems},
  author={Matei Zaharia and Omar Khattab and Lingjiao Chen and Jared Quincy Davis
          and Heather Miller and Chris Potts and James Zou and Michael Carbin
          and Jonathan Frankle and Naveen Rao and Ali Ghodsi},
  howpublished={\url{https://bair.berkeley.edu/blog/2024/02/18/compound-ai-systems/}},
  year={2024},
  note={Accessed: 2026-04-19}
}

@misc{gpt3-5-turbo,
  title        = {GPT-3.5 Turbo},
  author       = {{OpenAI}},
  year         = {2023},
  howpublished = {\url{https://platform.openai.com/docs/models/gpt-3-5-turbo}},
  note         = {Accessed: 2026-04-19}
}

@article{wang2024survey,
  title={A survey on large language model based autonomous agents},
  author={Wang, Lei and Ma, Chen and Feng, Xueyang and Zhang, Zeyu and Yang, Hao and Zhang, Jingsen and Chen, Zhiyuan and Tang, Jiakai and Chen, Xu and Lin, Yankai and others},
  journal={Frontiers of Computer Science},
  volume={18},
  number={6},
  pages={186345},
  year={2024},
  publisher={Springer}
}

@article{izacard2021unsupervised,
  title={Unsupervised dense information retrieval with contrastive learning},
  author={Izacard, Gautier and Caron, Mathilde and Hosseini, Lucas and Riedel, Sebastian and Bojanowski, Piotr and Joulin, Armand and Grave, Edouard},
  journal={arXiv preprint arXiv:2112.09118},
  year={2021}
}

@article{karthik2025bang,
  title={BANG: Billion-Scale Approximate Nearest Neighbour Search using a Single GPU},
  author={Karthik, V and Khan, Saim and Singh, Somesh and Simhadri, Harsha Vardhan and Vedurada, Jyothi},
  journal={IEEE Transactions on Big Data},
  year={2025},
  publisher={IEEE}
}

@inproceedings{joshi2017triviaqa,
  title={TriviaQA: A Large Scale Distantly Supervised Challenge Dataset for Reading Comprehension},
  author={Joshi, Mandar and Choi, Eunsol and Weld, Daniel S and Zettlemoyer, Luke},
  booktitle={Proceedings of the 55th Annual Meeting of the Association for Computational Linguistics (Volume 1: Long Papers)},
  pages={1601--1611},
  year={2017}
}

@article{kwiatkowski2019natural,
  title={Natural Questions: A Benchmark for Question Answering Research},
  author={Kwiatkowski, Tom and Palomaki, Jennimaria and Redfield, Olivia and Collins, Michael and Parikh, Ankur and Alberti, Chris and Epstein, Danielle and Polosukhin, Illia and Devlin, Jacob and Lee, Kenton and others},
  journal={Transactions of the Association for Computational Linguistics},
  volume={7},
  pages={452--466},
  year={2019}
}

@article{chen2021spann,
  title={Spann: Highly-efficient billion-scale approximate nearest neighbor search},
  author={Chen, Qi and Zhao, Bing and Wang, Haidong and Li, Mingqin and Liu, Chuanjie and Li, Zengzhong and Yang, Mao and Wang, Jingdong},
  journal={arXiv preprint arXiv:2111.08566},
  year={2021}
}

@article{jayaram2019diskann,
  title={Diskann: Fast accurate billion-point nearest neighbor search on a single node},
  author={Jayaram Subramanya, Suhas and Devvrit, Fnu and Simhadri, Harsha Vardhan and Krishnawamy, Ravishankar and Kadekodi, Rohan},
  journal={Advances in Neural Information Processing Systems},
  volume={32},
  year={2019}
}

@article{shaoscaling,
  title={Scaling retrieval-based language models with a trillion-token datastore},
  author={Shao, Rulin and He, Jacqueline and Asai, Akari and Shi, Weijia and Dettmers, Tim and Min, Sewon and Zettlemoyer, Luke and Koh, Pang W},
  journal={Advances in Neural Information Processing Systems},
  volume={37},
  pages={91260--91299},
  year={2024}
}

@inproceedings{ivf_original,
  author = {Sivic, Josef and Zisserman, Andrew},
  booktitle = {ICCV},
  pages = {1470-1477},
  publisher = {IEEE Computer Society},
  title = {Video Google: A Text Retrieval Approach to Object Matching in Videos.},
  year = 2003
}

@article{malkov_hnsw,
    author = {Malkov, Yu A. and Yashunin, D. A.},
    title = {Efficient and Robust Approximate Nearest Neighbor Search Using Hierarchical Navigable Small World Graphs},
    year = {2020},
    issue_date = {April 2020},
    publisher = {IEEE Computer Society},
    address = {USA},
    volume = {42},
    number = {4},
    issn = {0162-8828},
    url = {https://doi.org/10.1109/TPAMI.2018.2889473},
    journal = {IEEE Trans. Pattern Anal. Mach. Intell.},
}

@article{douze2024faiss,
  title={The faiss library},
  author={Douze, Matthijs and Guzhva, Alexandr and Deng, Chengqi and Johnson, Jeff and Szilvasy, Gergely and Mazar{\'e}, Pierre-Emmanuel and Lomeli, Maria and Hosseini, Lucas and J{\'e}gou, Herv{\'e}},
  journal={arXiv preprint arXiv:2401.08281},
  year={2024}
}

@article{lu2024turborag,
  title={TurboRAG: Accelerating Retrieval-Augmented Generation with Precomputed KV Caches for Chunked Text},
  author={Lu, Songshuo and Wang, Hua and Rong, Yutian and Chen, Zhi and Tang, Yaohua},
  journal={arXiv preprint arXiv:2410.07590},
  year={2024}
}

@article{xie2024ai,
  title={AI Metropolis: Scaling Large Language Model-based Multi-Agent Simulation with Out-of-order Execution},
  author={Xie, Zhiqiang and Kang, Hao and Sheng, Ying and Krishna, Tushar and Fatahalian, Kayvon and Kozyrakis, Christos},
  journal={arXiv preprint arXiv:2411.03519},
  year={2024}
}

@inproceedings{pytorch_2,
    author = {Ansel, Jason and Yang, Edward and He, Horace and Gimelshein, Natalia and Jain, Animesh and Voznesensky, Michael and Bao, Bin and Bell, Peter and Berard, David and Burovski, Evgeni and Chauhan, Geeta and Chourdia, Anjali and Constable, Will and Desmaison, Alban and DeVito, Zachary and Ellison, Elias and Feng, Will and Gong, Jiong and Gschwind, Michael and Hirsh, Brian and Huang, Sherlock and Kalambarkar, Kshiteej and Kirsch, Laurent and Lazos, Michael and Lezcano, Mario and Liang, Yanbo and Liang, Jason and Lu, Yinghai and Luk, C. K. and Maher, Bert and Pan, Yunjie and Puhrsch, Christian and Reso, Matthias and Saroufim, Mark and Siraichi, Marcos Yukio and Suk, Helen and Zhang, Shunting and Suo, Michael and Tillet, Phil and Zhao, Xu and Wang, Eikan and Zhou, Keren and Zou, Richard and Wang, Xiaodong and Mathews, Ajit and Wen, William and Chanan, Gregory and Wu, Peng and Chintala, Soumith},
    title = {PyTorch 2: Faster Machine Learning Through Dynamic Python Bytecode Transformation and Graph Compilation},
    year = {2024},
    address = {New York, NY, USA},
    url = {https://doi.org/10.1145/3620665.3640366},
    booktitle = {Proceedings of the 29th ACM International Conference on Architectural Support for Programming Languages and Operating Systems, Volume 2},
    pages = {929–947},
    numpages = {19},
    series = {ASPLOS '24}
}

@inproceedings{jiang2025rago,
  title={Rago: Systematic performance optimization for retrieval-augmented generation serving},
  author={Jiang, Wenqi and Subramanian, Suvinay and Graves, Cat and Alonso, Gustavo and Yazdanbakhsh, Amir and Dadu, Vidushi},
  booktitle={Proceedings of the 52nd Annual International Symposium on Computer Architecture},
  pages={974--989},
  year={2025}
}

@inproceedings{shen2025hermes,
  title={Hermes: Algorithm-system co-design for efficient retrieval-augmented generation at-scale},
  author={Shen, Michael and Umar, Muhammad and Maeng, Kiwan and Suh, G Edward and Gupta, Udit},
  booktitle={Proceedings of the 52nd Annual International Symposium on Computer Architecture},
  pages={958--973},
  year={2025}
}

@inproceedings{hu2025hedrarag,
  title={Hedrarag: Co-optimizing generation and retrieval for heterogeneous rag workflows},
  author={Hu, Zhengding and Murthy, Vibha and Pan, Zaifeng and Li, Wanlu and Fang, Xiaoyi and Ding, Yufei and Wang, Yuke},
  booktitle={Proceedings of the ACM SIGOPS 31st symposium on operating systems principles},
  pages={623--638},
  year={2025}
}

@inproceedings{sella2024flash,
  title     = {Flash is Driving Scale in RAG-Based LLMs},
  author    = {Sella, Assaf},
  booktitle = {Proceedings of the Flash Memory Summit (FMS)},
  year      = {2024},
  month     = {August},
  note      = {Session AIML-203-1},
  url       = {https://files.futurememorystorage.com/proceedings/2024/20240807_AIML-203-1_Sella.pdf}
}

@misc{zeromq,
  author       = {{The ZeroMQ authors}},
  title        = {{ZeroMQ}},
  howpublished = {\url{https://zeromq.org/}},
  year         = {2025}
}

@misc{milvus,
  author       = {{Milvus team}},
  title        = {{Milvus - The High-Performance
Vector Database Built for Scale}},
  howpublished = {\url{https://milvus.io/}},
  year         = {2025}
}

@misc{pgvector,
  author       = {{Pgvector team}},
  title        = {{Pgvector - Open-source vector similarity search for Postgres}},
  howpublished = {\url{https://github.com/pgvector/pgvector}},
  year         = {2025}
}

@misc{nonexhaustive_faiss,
  author       = {{FAISS team}},
  title        = {{FAISS index factory - Non-exhaustive search components}},
  howpublished = {\url{https://github.com/facebookresearch/faiss/wiki/The-index-factory#non-exhaustive-search-components}},
  year         = {2025},
  note         = {Accessed: 2026-04-19}
}

@inproceedings{indyk1998approximate,
  title={Approximate nearest neighbors: towards removing the curse of dimensionality},
  author={Indyk, Piotr and Motwani, Rajeev},
  booktitle={Proceedings of the thirtieth annual ACM symposium on Theory of computing},
  pages={604--613},
  year={1998}
}

@inproceedings{gionis1999similarity,
  title={Similarity search in high dimensions via hashing},
  author={Gionis, Aristides and Indyk, Piotr and Motwani, Rajeev and others},
  booktitle={Vldb},
  volume={99},
  number={6},
  pages={518--529},
  year={1999}
}

@inproceedings{datar2004locality,
  title={Locality-sensitive hashing scheme based on p-stable distributions},
  author={Datar, Mayur and Immorlica, Nicole and Indyk, Piotr and Mirrokni, Vahab S},
  booktitle={Proceedings of the twentieth annual symposium on Computational geometry},
  pages={253--262},
  year={2004}
}

@misc{nvidiagrace,
  author       = {{NVIDIA Corporation}},
  title        = {{NVIDIA Grace CPU Superchip}},
  year         = {2024},
  howpublished = {\url{https://www.nvidia.com/en-us/data-center/grace-cpu-superchip/}},
  note         = {Accessed: 2026-04-19}
}
\bibliographystyle{mlsys2026}

\appendix
\section{Artifact Appendix}

\subsection{Abstract}

This artifact contains the source code, evaluation scripts, and datasets for TeleRAG, an efficient retrieval-augmented generation (RAG) inference system that reduces latency and improves throughput via \emph{lookahead retrieval}, a prefetching mechanism that predicts required data and transfers them from CPU to GPU in parallel with LLM generation. The artifact reproduces Table~3 and Figures~9--14 of the paper, covering single-GPU latency (RTX4090), single-GPU throughput (H100), multi-GPU scaling (H200), and prefetch cluster hit rates. All experiments are automated using GNU Make and run within a Docker container to ensure reproducibility.

\subsection{Artifact check-list (meta-information)}

{\small
\begin{itemize}
  \item {\bf Algorithm:} IVF cluster-based lookahead retrieval with prefetching, greedy cache-aware batch scheduling.
  \item {\bf Program:} Python (\texttt{ragacc} library) and shell scripts.
  \item {\bf Compilation:} Docker build.
  \item {\bf Data set:} wiki\_dpr dataset \cite{karpukhin2020dense}, NaturalQuestions \cite{kwiatkowski2019natural}, HotpotQA \cite{yang2018hotpotqa}, and TriviaQA \cite{joshi2017triviaqa}.
  \item {\bf Run-time environment:} Docker (CUDA 12.8, Ubuntu 22.04, Python 3), NVIDIA Container Toolkit
  \item {\bf Hardware:} RTX~4090 (24\,GB) for single-GPU latency; H100 (80\,GB) for single-GPU throughput; up to 8$\times$H200 (140\,GB each) for multi-GPU scaling. 64+\,GB system RAM for single-GPU, 900+\,GB for 8-GPU experiments.
  \item {\bf Run-time state:} No other GPU-intensive processes should be running during evaluation (PCIe bandwidth contention affects results).
  \item {\bf Execution:} Automated via GNU Make targets.
  \item {\bf Metrics:} End-to-end latency (ms), throughput (queries/s), retrieval speedup, prefetch cluster hit rate (\%).
  \item {\bf Output:} CSV files, PDF plots, JSON hit rate results.
  \item {\bf How much disk space required (approximately)?:} $\sim$400\,GB (200\,GB dataset + models + index).
  \item {\bf How much time is needed to prepare workflow (approximately)?:} 1--2 hours (download models/dataset, build Docker image).
  \item {\bf How much time is needed to complete experiments (approximately)?:} up to 24 hours per GPU configuration.
  \item {\bf Publicly available?:} Yes.
  \item {\bf Code licenses (if publicly available)?:} Apache 2.0.
  \item {\bf Workflow framework used?:} GNU Make and Bash.
  \item {\bf Archived (provide DOI)?:} \url{https://doi.org/10.5281/zenodo.19361856}.
\end{itemize}
}

\subsection{Description}

\subsubsection{How delivered}

The artifact is available as a GitHub repository (\url{https://github.com/uw-syfi/TeleRAG}) that contains the \texttt{ragacc} library, evaluation scripts, and a Dockerfile. Models and datasets are downloaded separately from HuggingFace.\footnote{https://huggingface.co/datasets/lauyeeyu/TeleRAG-Dataset} Detailed setup instructions are provided in \texttt{docs/artifact-evaluation.md}.

\subsubsection{Hardware dependencies}

For Single-GPU experiments, 1$\times$ NVIDIA RTX~4090 or 1$\times$ H100 with 64+\,GB system RAM and PCIe Gen4/Gen5 is required. For Multi-GPU experiments, 8$\times$ NVIDIA H200 with 900+\,GB system RAM and PCIe Gen5 is required. See Table~\displayappendixref{2}{\ref{tab:hardware-setup}} for details.

\subsubsection{Software dependencies}

The artifact requires Docker with NVIDIA Container Toolkit. The Dockerfile installs all software dependencies. A Hugging Face account with access to gated Llama~3 and Mistral models is required to download models.

\subsubsection{Data sets}

The published dataset contains the vector index trained for the wiki\_dpr dataset \cite{karpukhin2020dense}, and pre-generated RAG pipeline traces for three datasets (NaturalQuestions, HotpotQA, and TriviaQA \cite{kwiatkowski2019natural,yang2018hotpotqa,joshi2017triviaqa}) with GPT-3.5-Turbo \cite{gpt3-5-turbo}.

\subsection{Installation}

Set up \sys by first cloning the TeleRAG repository.

\begin{verbatim}
$ git clone \
  https://github.com/uw-syfi/TeleRAG.git
\end{verbatim}

Download the required Hugging Face models and datasets as described in \texttt{docs/artifact-evaluation.md}. Some models, including Llama and Mistral, are gated and require accepting their licenses before download.

\begin{verbatim}
$ hf download <model name>
\end{verbatim}

Build the Docker image for \sys.

\begin{verbatim}
$ docker build -t telerag .
\end{verbatim}

Launch a detached GPU-enabled Docker container with the required capabilities and directory mounts for the repository, models, and dataset.

\begin{verbatim}
$ docker run --gpus all \
  --cap-add=SYS_NICE \
  -d -it --name telerag-ae \
  -v "$(pwd)":/app \
  -v <path to models>:/hf_models \
  -v <path to datasets>:/data/ \
  telerag:latest
\end{verbatim}

Then enter the running container.

\begin{verbatim}
$ docker exec -it telerag-ae /bin/bash
\end{verbatim}

Inside the Docker container, run the smoke test to verify the setup.

\begin{verbatim}
$ make smoke-test
\end{verbatim}




\subsection{Experiment workflow}

All experiments are orchestrated via GNU Make. Do not run experiments in parallel (\texttt{-j}) due to PCIe bandwidth contention.

\begin{enumerate}
  \item \textbf{Single-GPU latency (RTX~4090):} \texttt{make 4090} runs FAISS and TeleRAG on 6 RAG pipelines across 3 datasets with Llama-3.2-3B and Llama-3-8B. \texttt{make 4090-plots} generates Figure~\displayappendixref{9}{\ref{fig:result-speedup-rtx4090}}.
  \item \textbf{Single-GPU throughput (H100):} \texttt{make h100} runs hit rate calculation (Table~\displayappendixref{3}{\ref{tab:result-hitrate}}), batch evaluation with Llama-3-8B and Mistral-Small-22B. \texttt{make h100-plots} generates Figures~\displayappendixref{10}{\ref{fig:result-throughput-h100}} and~\displayappendixref{12}{\ref{fig:result-latency-breakdown-llama3-8b}}.
  \item \textbf{Multi-GPU scaling (H200):} \texttt{make h200} runs multi-GPU evaluation with 1--8 H200 GPUs plus scheduling ablations. \texttt{make h200-plots} generates Figures~\displayappendixref{11}{\ref{fig:result-throughput-multi-gpu}}, \displayappendixref{13}{\ref{fig:result-multi-gpu-cache-comparison}}, and~\displayappendixref{14}{\ref{fig:result-latency-breakdown-schedulers}}.
\end{enumerate}

Results are saved as CSV files in \texttt{evaluation/} and PDF plots in \texttt{figure/}.

\subsection{Evaluation and expected result}
The generated plots should be compared with the corresponding figures in the paper:

\begin{itemize}
  \item \textbf{Table~\displayappendixref{3}{\ref{tab:result-hitrate}}:} Prefetch cluster hit rates per pipeline on NQ.
  \item \textbf{Figure~\displayappendixref{9}{\ref{fig:result-speedup-rtx4090}}:} TeleRAG should show consistent latency speedup over FAISS across all 6 pipelines and 3 datasets on RTX~4090.
  \item \textbf{Figure~\displayappendixref{10}{\ref{fig:result-throughput-h100}}:} TeleRAG should achieve higher throughput than FAISS across batch sizes on H100.
  \item \textbf{Figure~\displayappendixref{11}{\ref{fig:result-throughput-multi-gpu}}:} Throughput should scale with GPU count (1--8 H200 GPUs) across all datasets.
  \item \textbf{Figure~\displayappendixref{12}{\ref{fig:result-latency-breakdown-llama3-8b}}:} Latency breakdown should show retrieval time reduction due to prefetching overlap.
  \item \textbf{Figure~\displayappendixref{13}{\ref{fig:result-multi-gpu-cache-comparison}}:} Cache-aware scheduling should improve throughput compared to without cache.
  \item \textbf{Figure~\displayappendixref{14}{\ref{fig:result-latency-breakdown-schedulers}}:} Scheduling overhead should be small relative to total latency.
\end{itemize}

Performance numbers may vary across runs due to system load, thermal throttling, and PCIe contention. Ensure no other GPU-intensive processes are running. If results differ significantly, restart the Docker container and re-run.

\subsection{Experiment customization}

Key parameters can be adjusted in the shell scripts under \texttt{artifact\_evaluation/}:

\begin{itemize}
  \item \texttt{N\_SAMPLES}: Number of evaluation samples (default: 1024 for single-GPU, 512 for multi-GPU).
  \item \texttt{N\_RUNS}: Number of benchmark runs (default: 5).
  \item \texttt{NPROBE}: Number of IVF clusters to probe (default: 256).
  \item \texttt{VM\_SIZE}: Prefetch buffer size in GB.
\end{itemize}

Prefetch budgets per pipeline/GPU/dataset are defined in \texttt{ragacc/pipeline\_budgets.py}.





\section{Background: Inverted File Index (IVF) for Vector Search}
\label{sec:ivf-math}

The \textit{vector index} is a crucial component of RAG applications that retrieves similar items from large datasets of high-dimensional vectors. 
Given the query vector $x \in \mathrm{R}^D$ and vector database $Y = \{y_0, \ldots, y_{N-1}\} \subset \mathrm{R}^D$, which comprises $N$ vectors, the vector index aims to find the $k$ nearest neighbors of $x$ from the database:
\begin{align*}
    \text{k-argmin}_{i = 0:N} d(x, y_i), \\
\end{align*}
where $D$ is the dimensionality of the vector determined by the embedding model, and $d$ denotes the distance function, which is typically the L2-norm or inner product~\cite{johnson2019billion}. 

To improve search efficiency, the inverted file index (IVF) algorithm is widely used for large-scale vector databases due to its simplicity and effectiveness. IVF partitions the datastore into \textit{clusters} and restricts searches to the most relevant clusters, reducing the computational cost.
To obtain the cluster assignment of each stored vector, it performs a clustering algorithm such as $k$-means~\cite{norouzi2013cartesian} and partitions the database into $N_c$ clusters:
\begin{align*}
    \{C_1, C_2, \ldots, C_{N_c}\} = k\text{-means}(Y, N_c),
\end{align*}
where $C_j$ is the set of vectors assigned to the $j$-th cluster, and the cluster centroids $\{c_1, c_2, \ldots, c_{N_c}\}$ are obtained by taking the mean of vectors in $\{C_1, C_2, \ldots, C_{N_c}\}$.
Then, each database vector $y_i \in Y$ is assigned to the nearest cluster center:
\begin{align*}
    \text{Cluster}(y_i) = \text{argmin}_{j=1:N_c} d(y_i, c_j).
\end{align*}

With the trained centroids and cluster assignment, we can perform a more efficient index search. There are two steps involved in the IVF index search. First, it will identify the nearest $L$ cluster centroids of a query vector $x$:
\begin{align*}
    \{c_{j_1}, c_{j_2}, \ldots, c_{j_L}\} = L\text{-argmin}_{j=1:C} \ d(x, c_j).
\end{align*}
This step is also referred to as the coarse-grained search in IVF, as it identifies candidates at the cluster level.
Second, the fine-grained search is performed in the $L$ nearest clusters and identifies the $k$ closest vectors of $x$:
\begin{align*}
    k\text{-argmin}_{y_i \in \cup_{l=1}^L C_{j_l}} d(x, y_i).
\end{align*}
This involves sorting.
In this way, IVF reduces search space and accelerates the retrieval process.
Here, the hyperparameter $L$ from the first step is also referred to as \nprobe \cite{douze2024faiss}.
When \nprobe is larger, the IVF retrieval will be more accurate as it searches more clusters.
However, the retrieval latency is longer as more computation and data are needed.

\section{Finding Prefetching Amount}
\label{sec:math-analysis-prefetch-amount}

A key challenge in \sys is to balance the benefit of reducing retrieval latency by prefetching data against the overhead of CPU-GPU transfers. Prefetching more clusters reduces the subsequent retrieval time but can also extend the transfer phase; if it extends beyond the LLM generation window, we lose the advantage of overlap and potentially introduce additional delay. Still, additional delays in data transfer may be worthwhile if they substantially reduce retrieval latency. 
To guide the choice of the optimal amount of data to prefetch, we develop a mathematical model based on profile information.

\pgheading{Mathematical model}
Here, we denote $b_p$ as the number of bytes to prefetch and $B$ as the CPU-GPU bandwidth.
The optimal amount of data to prefetch is denoted as $b_p^*$.
To start, we let $t_{1}$ represent the combined time of prefetching and pre-retrieval LLM generation, and $t_{2}$ represent the retrieval time.
We have: $t_1 = \max(t_{\text{LLM}}, t_{p})$ and $t_2 = \max(t_{c}, t_{g})$,
where $t_{\text{LLM}}$ is the time for LLM generation, $t_{p}$ is prefetching time, $t_{c}$ is the CPU retrieval time, and $t_{g}$ is the GPU retrieval time. The objective is to minimize $t_1 + t_2$.

Since prefetching time $t_{p}$ is proportional to $b_p$, we express $t_1$ as a piecewise function:
\begin{equation}\label{equation:t1}
t_1 = \begin{cases}
    t_{\text{LLM}}, & \text{if } b_p \le B \cdot t_{\text{LLM}}, \\[6pt]
    \frac{b_p}{B}, & \text{if } b_p > B \cdot t_{\text{LLM}},
\end{cases}
\end{equation}
From Eq.~\ref{equation:t1}, if we prefetch fewer bytes than can be transferred during LLM generation, $t_1$ is effectively just $t_{\text{LLM}}$ because the transfer overlaps completely with LLM execution.

As shown in the analysis section, GPU retrieval ($t_g$) is generally much faster than CPU retrieval ($t_c$). Thus, we assume $t_c \gg t_g$. 
As CPU has a limited parallelism, $t_c$ usually grows proportionally with the number of clusters to be processed~\cite{jiang2023co}:
\begin{align}
t_2 = t_c = r_{\mathrm{miss}} \times n_{\mathrm{probe}} \times t_{cc},
\end{align}
where $r_{\mathrm{miss}}$ is the miss rate (the percentage of IVF clusters that are not cached on the GPU), $n_{\mathrm{probe}}$ is the total number of clusters to search, and $t_{cc}$ is the CPU time to search a single cluster.
Increasing $b_p$ can only decrease or maintain the current miss rate, \ie $\frac{\mathrm{d} r_{\mathrm{miss}}}{\mathrm{d} b_p} \le 0$.
Moreover, because clusters are prefetched in order of descending likelihood, we assume $r_{\mathrm{miss}}$ is either linear or a concave up function\footnote{An upward U-shaped function whose second derivative is positive.} of $b_p$, \ie $\frac{\mathrm{d}^2 r_{\mathrm{miss}}}{\mathrm{d} b_p^2} \ge 0$.

We now examine two cases:
\begin{itemize}
\item \textbf{Case 1:} $b_p^* \le B \cdot t_{\text{LLM}}$.
Here, $t_1 = t_{\text{LLM}}$ is constant because prefetching is fully overlapped with LLM generation. Since increasing $b_p$ in this regime will not increase $t_1$ and cannot worsen the miss rate, pushing $b_p$ to the boundary $B \cdot t_{\text{LLM}}$ minimizes $t_1 + t_2$. Hence,
\begin{equation}\label{equation:minimum-1}
b_p^* = B \cdot t_{\text{LLM}}.
\end{equation}

\item \textbf{Case 2:} $b_p^* > B \cdot t_{\text{LLM}}$.
In this region, $t_1$ grows linearly with $b_p$, and we have:
$\frac{\mathrm d^2}{\mathrm d b_p^2}(t_1 + t_2) = \frac{\mathrm{d}^2 r_{\mathrm{miss}}}{\mathrm{d} b_p^2} \cdot n_{\mathrm{probe}} \cdot t_{cc} \geq 0.$
Therefore, $t_1 + t_2$ is concave up, allowing at most one minimum. At the minimum point, we have:
\begin{equation}
\frac{\mathrm d}{\mathrm d b_p} (t_1 + t_2) = 0 \implies \frac{1}{B} + \frac{\mathrm{d} r_{\mathrm{miss}}}{\mathrm d b_p} \cdot n_{\mathrm{probe}} \cdot t_{cc} = 0.
\end{equation}
From this, we obtain:
\begin{equation}
\label{equation:minimum-2}
b_p^*
= B \times n_{\mathrm{probe}} \times t_{cc} \times \Delta r_{\mathrm{miss}},
\end{equation}
where $\Delta r_{\mathrm{miss}}$ is the decrement of the miss rate for this round.
If $b_p^*$ is indeed larger than $B \cdot t_{\text{LLM}}$, it becomes the global minimum; otherwise, the solution reverts to Case~1.
\end{itemize}

In summary, our analysis shows that the optimal prefetch amount $b_p^*$ can only lie at one of two points:
(1)~Prefetch until LLM generation completes (\ie $b_p = B \cdot t_{\text{LLM}}$).
(2)~A point determined by Eq.~\ref{equation:minimum-2}.
However, under typical CPU-GPU bandwidth (\eg 64\,GB/s on PCIe\,5), 
the time spent loading additional clusters often outweighs any retrieval latency reduction from lowering \(r_{\mathrm{miss}}\).
Consequently, the second scenario in Eq.~\ref{equation:minimum-2} becomes nearly infeasible in practice. 
Therefore, on current hardware, \emph{prefetching exactly until LLM execution ends} is generally the most effective choice.

\section{Implementation Details}
\label{sec:impl}

\sys is implemented in Python and also leverages PyTorch's~\cite{pytorch_2} operator ecosystem for efficient computation. 
The datastore index is initially constructed using FAISS~\cite{douze2024faiss}, and its data structures, such as IVF centroids and cluster data, are converted to PyTorch tensors.

At runtime, cluster data is loaded into a contiguous pinned memory region on the CPU, enabling non-blocking memory copies to the GPU. A fixed-size contiguous buffer on the GPU is allocated based on the user's configuration or GPU memory capacity during runtime.

We use separate processes for the schedulers and services in \sys{} to avoid bottlenecks in Python global interpreter lock.
For the communication between the scheduler and the services,
we also leverage ZeroMQ~\cite{zeromq} as the communication framework.
The message is encoded and decoded with Pickle in Python.
For the LLM service, we use SGLang~\cite{zheng2024_sglang} as the
LLM backend.
For the retrieval service, to implement the index search with GPU-CPU cooperation, on GPU, we use a single matrix-vector multiplication that computes distances for all prefetched vectors; on CPU, we utilize \texttt{multithreading} in Python to parallelize similarity searches across clusters.
We then move the distances computed from CPU to GPU, merge with results on GPU and perform global sorting on GPU.

To monitor the hotness of cached clusters better, we use a
\textbf{decay-based} approach to track the hotness of cached clusters.
Every cached cluster has a hotness value.
Upon fetching to the GPU, the cluster is essentially cached
and will be assigned an initial value $h_0 = h_{\text{init}}$.
Then, when a new round of hotness update starts,
the hotness will be divided by a decay factor $d$,
and if the cluster is used in the previous round,
a hotness update value $h_{\text{inc}}$ will be added to the hotness.
Therefore, the transition of hotness from the round $r$ to round $r+1$
can be given as:
\begin{equation}
h_{r+1} = \begin{cases}
h_r / d,\text{ if not used in round $r$,}\\
h_r / d + h_{\text{inc}},\text{ if used in round $r$.}
\end{cases}
\end{equation}
In this way, the process can be parallelized easily on GPU, minimizing the
overhead of hotness tracking in terms of both the time and computation.
We also set a cache fraction, which means if the cache exceeds the
fraction, we need to evict some clusters.
The cache fraction is set to $0.5$ by default.
To make evaluation results stable and reproducible, we do cache
eviction for the excessive clusters and consolidate the GPU memory
after serving each batch.
This makes sure that we will not over-evaluate the performance of the
caching system.

Both schedulers use greedy search to minimize overhead. The prefetching scheduler groups similar queries into micro-batches, and feeds them into the cache-aware scheduler to assign micro-batches to GPUs. The cache-aware scheduler will get the number of overlapped clusters on each GPU for each micro-batch, and assign the micro-batch to the GPU with the most overlapped clusters.

\section{Detailed Index Configurations}
\label{sec:detailed-index-config}

\begin{table}[htb]
    \centering 
    \resizebox{0.98\columnwidth}{!}{
    \begin{tabular}{lc}
        \toprule
        Specification & Value \\
        \midrule
        Dataset & wiki\_dpr \cite{karpukhin2020dense} \\
        Dataset size & 2.1 billion tokens \\
        $\#$ of chunks & 21 million \\
        $\#$ of IVF cluster & 4096 \\
        Embed model & Contriever \cite{izacard2021unsupervised} \\
        Embed dimension & 768 \\
        Index type & \texttt{FLAT}\tablefootnote{Original embedding without compression for the best retrieval precision.} \\
        Distance metric & \texttt{Inner Product}\\ 
    Index size & 61\,GB \\
        \bottomrule
    \end{tabular}
    }
    \caption{Detailed configurations of our retrieval index.}
\label{tab:index-setup}
\end{table}

Table~\ref{tab:index-setup} shows the detailed configurations of our retrieval index used in the evaluation.

\section{Additional Evaluation Results}

\begin{figure}[tb]
\centering
    \begin{subfigure}[t]{.8\columnwidth}
    \includegraphics[width=\columnwidth]{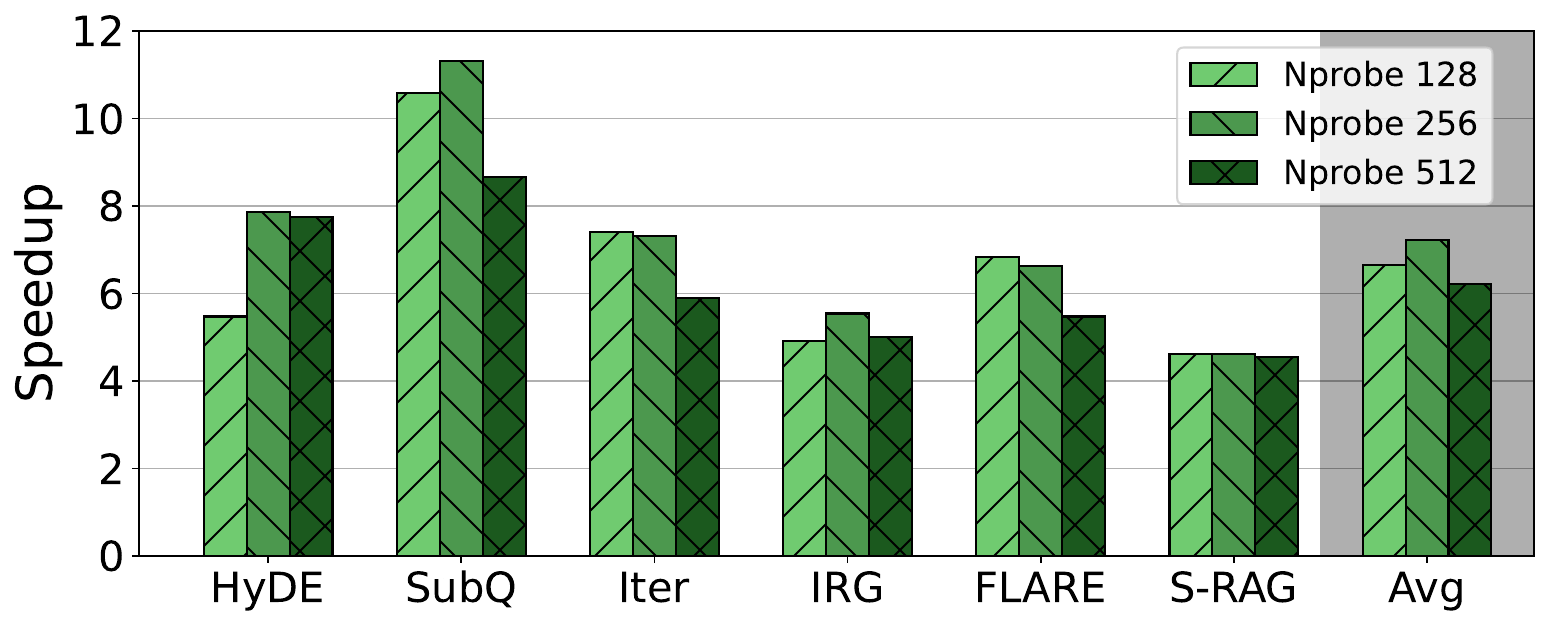}
    \caption{\llamasmall with an \gpusmall GPU.}
    \end{subfigure}
    
    \begin{subfigure}[t]{.8\columnwidth}
    \includegraphics[width=\columnwidth]{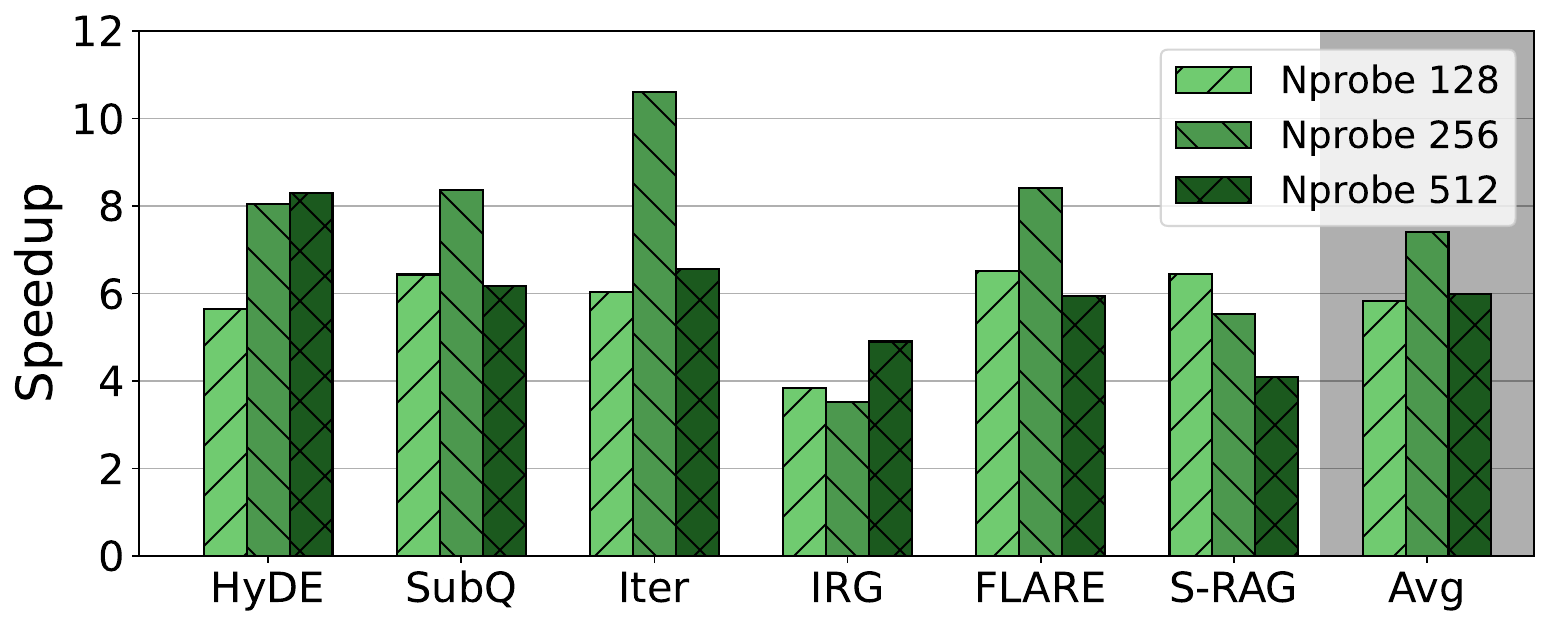}
    \caption{\llama with a \gpularge GPU.}
    \end{subfigure}
    \caption{Single-query retrieval speedup on NQ with different \nprobe{} values.}
    \label{fig:retrieval-speedup}
\end{figure}

\subsection{Retrieval Speedups Across \nprobe{}}

Figure~\ref{fig:retrieval-speedup} shows the retrieval latency reduction on NQ.
We observe consistent speedups for all \nprobe values.
The greatest speedups are achieved at \nprobe 256, with average speedups of $7.21\times$ and $7.41\times$ on \gpusmall and \gpularge, respectively.
As a larger \nprobe value is used, the retrieval performance of \sys becomes constrained by missed clusters on the CPU, given our fixed prefetch budget across varying \nprobe values.
However, RAG inference with higher \nprobe will result in longer retrieval latency, and hence \sys still has a significant latency improvement over the CPU retrieval baseline.


\end{document}